\documentclass[prd,aps,nofootinbib,preprintnumbers,showpacs,showkeys]
{revtex4}
\usepackage{graphicx,epsf,amsfonts,amssymb,amsbsy}
\newcommand{\ds}{\displaystyle}
\newcommand{\vev}[1]{\langle#1\rangle}
\newcommand{\mat}{\left ( \begin{array}}
\newcommand{\emat}{\end{array} \right )}
\newcommand{\vect}{\left ( \begin{array}{c}}
\newcommand{\evect}{\end{array} \right )}

\preprint{HU-EP-06/25}
\begin{document}

\title{ \bf Mesons and diquarks in neutral color
superconducting quark matter with $\beta$-equilibrium}
\author{D.~Ebert$^1$, K. G.~Klimenko$^2$, and V. L.
Yudichev$^3$}
\affiliation{$^1$ Institut f\"ur Physik,
Humboldt-Universit\"at zu Berlin, 10115 Berlin, Germany}
\affiliation{$^2$ Institute of High Energy Physics, 142281,
Protvino, Moscow Region, Russia}
\affiliation{$^3$ Joint Institute for Nuclear Research, 141980,
Dubna, Moscow Region, Russia}

\begin{abstract}
The spectrum of  meson and diquark excitations in cold
color-superconducting (2SC) quark matter is investigated under
local color and electric neutrality constraints with
$\beta$-equilibrium. A 2-flavored Nambu--Jona-Lasinio type model
including a baryon $\mu_B$, color $\mu_8$, and electric  $\mu_Q$
chemical potentials is used. The contribution from free electrons
to the free energy is added to take into account the
$\beta$-equilibrium. The sensitivity of the model to the tuning of
the interaction constants in the diquark ($H$) and quark--antiquark
($G$) channels is examined for two different parameterization schemes
by choosing the ratio $H/G$ to be 3/4 and 1, respectively. At
$H=3G/4$ the gapless- and at $H=G$ the gapped neutral color
superconductivity is realized. It is shown that color
and electrical neutrality together with $\beta$-equilibrium lead
to a strong mass splitting within the pion isotriplet in
the 2SC phase (both gapped and gapless), in contrast with
non--neutral matter. The $\pi$- and $\sigma$-meson masses are
evaluated to be $\sim300$ MeV. It is also shown that the
properties of the physical $SU(2)_c$-singlet diquark excitation in
the 2SC ground state varies for different parameterization
schemes. Thus, for $H=3G/4$ one finds  a heavy resonance with mass
$\sim$ 1100 MeV in the non--neutral (gapped) case, whereas, if
neutrality is imposed, a stable diquark with mass
$\sim |\mu_Q|\sim$ 200 MeV appears in the gapless 2SC environment.
For a stronger attraction in the diquark channel ($H=G$), there is
again a resonance (with the mass $\sim$ 300 MeV) in the neutral
gapped 2SC phase. Hence, the existence of the stable massive
SU(2)$_c$-singlet diquark excitation is a new peculiarity of the
gapless 2SC. In addition, the behaviour of the diquark mass in
vacuum, i.~e.\ , at $\mu_B=0$, as a function of $H$ has been
investigated.
\end{abstract}


\maketitle
\maketitle

\section{Introduction}

According to  modern theoretical observations made in the
framework of perturbative QCD, at asymptotically high baryonic
densities and low temperatures the strongly interacting quark
matter is expected to undergo a phase transition to the color
superconducting state \cite{love,son}. Unfortunately, a
perturbative QCD analysis is not applicable at moderate baryon
densities (which might exist inside compact stars or in heavy ion
collision experiments) and a study is usually done with the help
of effective theories, such as the Nambu--Jona-Lasinio (NJL) model
\cite{njl,eguchi,ebvolkov}.

In spite of the lack of quark confinement in NJL, it successfully
describes low-energy pseudoscalar and vector mesons in the
hadronic phase (see, e.~g., \cite{ebrein,hatsuda}). This success is
provided by the fact that many of light meson properties, e.~g.,
masses, are driven by chiral symmetry, rather than by confinement.
Moreover, the chiral phase transition at high temperatures and/or
density expected in QCD, is naturally described by the NJL model
\cite{hatsuda,bernard}. The NJL model is also well-suited
 for the consideration of a hot and/or  dense medium under the
influence of external conditions \cite{odintsov,incera} as well as
for the investigation of different physical processes in it
\cite{radzhabov,ebert}.

In the earlier studies of color superconductivity
\cite{skp,alford,yudichev} for the case of two-flavor quark matter
($u$ and $d$ quarks),  only the influence of baryonic density  was
taken into account. From these investigations, it became evident
that the two-flavor color superconducting phase (2SC)
might be yet present at rather small values of $\mu_B\sim 1$~GeV,
i.~e., at baryon densities only several times larger than the
density of ordinary nuclear matter (see reviews
\cite{alford,buballa,hs}). This is just the density of compact
star cores.

The quark matter inside compact stars is considered to be
electrically and color neutral in a bulk. Moreover, there must be an
equilibrium between the gain and loss in the $\beta$-decay  $d\to
u+e+\bar\nu_e$ (here $e$ is the electron and $\nu_e$ is the electron
neutrino), the so-called $\beta$-equilibrium. All these physical
constraints must be taken into account when studying the  equation of
states for a compact star. To do this in the NJL model, additional
chemical potentials related to the electric charge density of quarks
and electrons as well as to color charges must be introduced.
Recently, an intensive theoretical
study of neutral color superconducting quark matter has been given
(see, e.~g.,
\cite{buballa,hs,ar,sh2,zhuang,bs,he,abuki,h,g,hashimoto}),
which revealed a new possible ground state of the 2SC phase, where
some additional number of quasiparticles with a gapless dispersion
law is appeared \cite{sh2} (it is the so-called gapless color
superconductivity g2SC, an antipode to the usual, gapped 2SC).

There is a great interest in the study of different excitations of
the 2SC phase, as the application of its results in  some related
investigations, e.~g.\ in astrophysics and heavy-ion collision
experiments it may help to reveal some observable effects evidencing
the formation of a quark-gluon plasma. In particular, the bosonic
excitations of the 2SC phase ground state, such as $\pi$- and
$\sigma$-mesons as well as diquarks, are expected to be copiously
produced in dense medium with rather strong correlations between
quarks and antiquarks (just this thing is realized at moderate
baryonic densities) and affect some scattering and decay processes
to a visible effect. Diquarks on their own are also of great
interest in hadron physics because of their  importance in
determining baryon properties \cite{lawley}. As to the compact
stars, one may find an influence of these particles on the
equation of state and on the cooling process. Moreover, in dense
matter, the deconfinement phase transition might be accompanied by
the appearance of Bose--Einstein condensed diquark matter
\cite{pirner} etc.

In our recent papers \cite{bekvy,eky1}, we have studied the masses
of mesons and diquarks that are formed in cold ($T=0$) and dense
quark matter in the framework of a 2-flavored NJL model with
baryon chemical potential $\mu_B$. In particular, it was shown
that in the 2SC phase the meson masses lie in the interval
330$\div$500 MeV, depending on the values of $\mu_B\in
(1050,1200)$~MeV. Since the original SU(3)$_c$ color symmetry of
the model is spontaneously broken down in this phase to SU(2)$_c$,
one may expect the appearance of five Nambu--Goldstone bosons.
However, we have proved that the abnormal number of three, instead
of five, massless bosons is allowed for the diquark sector of the
model. In addition, there are two light diquarks as well as a
heavy diquark resonance that is an SU(2)$_c$-singlet with the mass
$\sim 1100$~MeV. Qualitatively, the local color neutrality
constraint does not affect the masses of mesons and the
SU(2)$_c$-singlet diquark. However, in this case only one
Nambu--Goldstone boson and four light diquarks are present in
the 2SC phase in the NJL model \cite{eky2} (see also the discussion
at the end of the paper). At nonzero temperature some of the
properties of mesons and diquaks in a strongly interacting quark
matter were discussed in \cite{hejz,Kitazawa}.

In the present paper, we continue our investigation of mass
spectra for mesons and diquarks, imposing the local electrical
neutrality and $\beta$-equilibrium to the cold 2SC medium, in
addition to the color neutrality. As in our previous papers, we
use a 2-flavored NJL, where additional color ($\mu_8$) and
electric ($\mu_Q$) chemical potentials are introduced. It will be
shown that in the 2SC phase, the color neutrality constraint
supplemented by the electrical neutrality and $\beta$-equilibrium
drastically changes the mass spectrum of the $\pi$-, $\sigma$- and
diquark, in comparison with non--neutral quark matter.

The paper is organized as follows. In Section II, the
thermodynamic potential as well as the effective action of the NJL
model, extended with baryon ($\mu_B$), color ($\mu_8$), and
electric ($\mu_Q$) chemical potentials, are obtained in the
one-loop approximation in $\beta$-equilibrium. Further, in Section
III, the gap equations and the phase diagram of quark matter are
investigated under the local color and electrical neutrality
constraints. Here, the behaviour of $\mu_8$ and $\mu_Q$ vs $\mu_B$
are obtained for  neutral quark matter with 2SC type color
superconductivity within two different parameterization schemes:
$H=3G/4$ and $H=G$. In the first case, $H=3G/4$, a gapless 2SC phase
revealed itself, whereas for $H=G$ a gapped phase is preferred. In
Sections IV and V, some peculiarities of the mass spectra of the
$\pi$, $\sigma$ mesons and scalar diquarks are investigated both in
the gapless and gapped neutral 2SC phases. (In addition, the 
influence of the diquark channel coupling constant on the diquark
mass in the vacuum, i.~e. at $\mu_B=0$, is also considered.) Finally,
in the Appendix, the expression for the quark propagator in the
Nambu--Gorkov representation  is obtained.

\section{The model and the effective action}

Our investigation is based on the NJL type model with two quark
flavors. Its Lagrangian describes the interaction in the 
quark--antiquark as well as scalar diquark channels:
\begin{eqnarray}
 L_q=\bar q\Big [\gamma^\nu i\partial_\nu-
m\Big ]q+ G\Big [(\bar qq)^2+
(\bar qi\gamma^5\vec\tau q)^2\Big ]+H\sum_{A=2,5,7}
[\bar q^Ci\gamma^5\tau_2\lambda_{A}q]
[\bar qi\gamma^5\tau_2\lambda_{A} q^C],
\label{1}
\end{eqnarray}
where the quark field $q\equiv q_{i\alpha}$ is a flavor doublet
($i=1,2$ or $i=u,d$) and color triplet ($\alpha=1,2,3$ or
$\alpha=r,g,b$) as well as a four-component Dirac spinor;
$q^C=C\bar q^t$ and $\bar q^C=q^t C$ are charge-conjugated
spinors, and $C=i\gamma^2\gamma^0$ is the charge conjugation
matrix (the symbol $t$ denotes the transposition operation). It is
supposed that up and down quarks have an equal current (bare) mass
$m$. Furthermore,  $\tau_a$ stands for Pauli matrices, and
$\lambda_A$ for Gell-Mann matrices in flavor and color space,
respectively. Clearly, the Lagrangian $L_q$ is invariant under
transformations from color SU(3)$_c$ as well as baryon U(1)$_B$
groups. In addition, at $m=0$ this Lagrangian is invariant under
the chiral SU(2)$_L\times$SU(2)$_R$ group. At $m\ne 0$ the chiral
symmetry is broken to the diagonal isospin subgroup  SU(2)$_I$
with the generators $I_k=\tau_k/2$ ($k=1,2,3$). Moreover, in our
system the electric charge is conserved, too, since $Q=I_3+B/2$,
where $I_3$ is the third generator of the isospin group SU(2)$_I$,
$Q$ is the electric charge generator, and $B$ is the baryon charge
generator (evidently,  these quantities are unit matrices in color
space, but in flavor space they are $Q=diag(2/3,-1/3)$,
$I_3=diag(1/2,-1/2)$ and $B=diag(1/3,1/3)$). If the Lagrangian
(\ref{1}) is obtained from the QCD one-gluon exchange
approximation, then $H=3G/4$. In addition, we find it interesting
to deal with another relation between coupling constants, $H=G$,
which results in qualitatively different model properties (see
below).

In order to take into account $\beta$-equilibrium, we  include
electrons into our consideration, extending the
Lagrangian as follows
\begin{eqnarray}
L_{qe}=L_q+\bar e \gamma^\nu i\partial_\nu e.
  \label{2}
\end{eqnarray}
Here $e$ is the electron spinor field (for simplicity, electrons are
taken to be massless). Clearly, the Lagrangian (\ref{2})
is well-suited for the description of different processes in the
vacuum, i.~e.\ in the empty space. Since the prime object of the
present paper is the consideration of dense medium properties, we
extend the Lagrangian (\ref{2}) by including terms with charge
densities and chemical potentials as it is usually done in
statistical physics
\begin{eqnarray}
L=L_{qe}+\mu_BN_B+\mu_QN_Q+\mu_8N_8.
  \label{3}
\end{eqnarray}
 In (\ref{3}), $N_B$, $N_Q$, $N_8$ are baryon,
electric and 8\textit{th}-color charge
density expressions, respectively;  $\mu_B$, $\mu_Q$, $\mu_8$ are the
corresponding chemical potentials.
Recall that
\begin{eqnarray}
N_B=\bar qB\gamma^0q,~~~~N_Q=\bar qQ\gamma^0q-\bar
e\gamma^0e,~~~~N_8=\bar qT_8\gamma^0q,
  \label{4}
\end{eqnarray}
where $T_8\equiv\sqrt{3}\lambda_8=diag(1,1,-2)$ is a matrix in the
color space. From equations (\ref{4}), we have
\begin{eqnarray}
\mu_BN_B+\mu_QN_Q+\mu_8N_8=\mu_e\bar e\gamma^0e+
\sum_{i,\alpha}\mu_{i\alpha}\bar q_{i\alpha}\gamma^0q_{i\alpha}\equiv
\mu_e\bar e\gamma^0e+\bar q\hat\mu\gamma^0q,
  \label{5}
\end{eqnarray}
where $\mu_e$ is the electron number chemical potential, and
$\mu_{i\alpha}$ is the chemical potential for the number of quarks
with color $\alpha$ and flavor $i$. Obviously, one has
\begin{eqnarray}
&& \mu_{ur}=\mu_{ug}=\frac {\mu_B}3+\frac
{2\mu_Q}3+\mu_8,~~~~\mu_{dr}=\mu_{dg}=\frac {\mu_B}3-\frac
{\mu_Q}3+\mu_8,\nonumber\\
&&\mu_{ub}=\frac {\mu_B}3+\frac {2\mu_Q}3-2\mu_8,~~~~\mu_{db}=\frac
{\mu_B}3-\frac {\mu_Q}3-2\mu_8,~~~~\mu_e=-\mu_Q,  \label{6}\\
&& ~~~~~~~~~~~\hat\mu=\mu_B/3+\mu_QQ+\mu_8T_8=
\tilde\mu+\delta\mu\tau_3+\mu_8T_8,
 \label{7}
\end{eqnarray}
where the last equality in (\ref{7}) is obtained due to the above
mentioned relation $Q=I_3+B/2$; moreover $\tilde\mu=\mu_B/3+\mu_Q/6$,
$\delta\mu =\mu_Q/2$. It follows from (\ref{6}) that
$\mu_{d\alpha}=\mu_{u\alpha}+\mu_{e}$ for each color $\alpha$. The
matrix $\hat\mu$ is  diagonal in
the six-dimensional (color)$\times$(flavor) space, and its  matrix
elements are just the quantities $\mu_{i\alpha}$ from (\ref{6}).

If all chemical potentials in (\ref{3}) are nonzero and
independent quantities, then SU(3)$_c$ and SU(2)$_I$ are not the
symmetry groups of this Lagrangian. Instead, due to the $\mu_8$- and
$\mu_Q$-terms, it is symmetric under the reduced color
SU(2)$_c\times$U(1)$_{\lambda_8}$ and flavor U(1)$_{I_3}$ groups.
With the local neutrality imposed, the chemical potentials
in (\ref{3}) are, however, no more independent quantities of the
model. Equating further $\vev{N_Q}$ and $\vev{N_8}$ to zero,
the chemical potentials are subjected to two constraints, thereby
fixing two of them. As a result, $\mu_Q$ and $\mu_8$ become dependent
on $\mu_B$. It turns out (see below) that there exists a critical
value of the baryon chemical potential $\mu_B^c$ in the locally
neutral matter which separates two phases: if $\mu_B<\mu_B^c$ the
normal quark matter phase with $\mu_Q =0$ and $\mu_8 =0$ is formed,
and the Lagrangian (\ref{3}) is an SU(3)$_c\times$SU(2)$_I$ invariant
one; if $\mu_B>\mu_B^c$ (2SC phase), the $\mu_Q$ and $\mu_8$ are not
already equal to zero and have a nontrivial $\mu_B$-dependence. It
means that at $\mu_B=\mu_B^c$ the SU(3)$_c\times$SU(2)$_I$ symmetry
of the Lagrangian (\ref{3}) is explicitly (not spontaneously) broken
by the chemical potential terms (i.~e.\ no Nambu--Goldstone bosons
must appear) to the color SU(2)$_c\times$U(1)$_{\lambda_8}$ and
flavor U(1)$_{I_3}$ groups. (Note that the color U(1)$_{\lambda_8}$
group is broken spontaneously in the 2SC phase, see below.) Due to
the above--mentioned flavor symmetry transformation in the critical
point $\mu_B^c$, one could expect that  all pions would have equal
masses for $\mu_B<\mu_B^c$, whereas at larger  $\mu_B$ ($\mu_B>
\mu_B^c$) the pion mass splitting should occur in neutral matter.

To study the phase diagram of the system and
the mass spectra of meson and diquark excitations, we need to get the
thermodynamic potential as well as an effective action up to second
order for the bosonic degrees of freedom. Since electrons and quarks
are not mixing, the total thermodynamic potential $\Omega$ of the
system is the sum of its electronic $\Omega_e$ and quark
$\Omega_q$ parts:  $\Omega =\Omega_q+\Omega_e$. It is well-known that
$\Omega_e=-\mu^4_e/12\pi^2$. To obtain $\Omega_q$, we start from the
Lagrangian describing the  quark contribution only
(see (\ref{3})),
\begin{eqnarray}
{\cal L}=L_q+\bar q\hat\mu\gamma^0q,
  \label{8}
\end{eqnarray}
where $\hat\mu$ is the quark number chemical potential matrix,
defined in (\ref{7}). The linearized version of Lagrangian (\ref{8})
that contains auxiliary bosonic fields has the following form
\begin{eqnarray}
\tilde {\cal L}\ds &=&\bar q\Big [\gamma^\nu i\partial_\nu
+\hat\mu\gamma^0
 -\sigma - m -i\gamma^5\pi_a\tau_a\Big ]q
 -\frac{1}{4G}\Big [\sigma\sigma+\pi_a\pi_a\Big ]
 \nonumber\\ &-&\frac1{4H}\Delta^{*}_{A}\Delta_{A}-
 \frac{\Delta^{*}_{A}}{2}[\bar q^Ci\gamma^5\tau_2\lambda_{A} q]
-\frac{\Delta_{A}}{2}[\bar q i\gamma^5\tau_2\lambda_{A}q^C],
\label{9}
\end{eqnarray}
where here and later, a summation over  repeated
indices $a=1,2,3$ and $A,A'=2,5,7$ is implied.
Clearly, the Lagrangians (\ref{8}) and (\ref{9}) are equivalent, as
can be seen by using the equations of motion for bosonic fields,
which take the form 
\begin{eqnarray}
\sigma (x)=-2G(\bar qq),~~\pi_a(x)=-2G(\bar qi\gamma^5\tau_a q),~~
\Delta_{A}(x)\!\!&=&\!\!-2H(\bar
q^Ci\gamma^5\tau_2\lambda_{A}q),~~
\Delta^{*}_{A}(x)=-2H(\bar qi\gamma^5\tau_2\lambda_{A} q^C).
\label{10}
\end{eqnarray}
One can easily see from (\ref{10}) that the mesonic fields
$\sigma(x),\pi_a(x)$ are real quantities, i.~e.\
$(\sigma(x))^\dagger=\sigma(x),~~
(\pi_a(x))^\dagger=\pi_a(x)$ (the superscript symbol $\dagger$
denotes the hermitian conjugation), but all diquark fields
$\Delta_{A}(x)$ are complex scalars, so
$(\Delta_{A}(x))^\dagger=\Delta^{*}_{A}(x)$.
Clearly, the real $\sigma(x)$ and $\pi_a(x)$ fields
are color singlets, whereas scalar diquarks
$\Delta_{A}(x)$ form a color antitriplet $\bar 3_c$ of the
SU(3)$_c$ group. If some of the scalar diquark fields have a nonzero
ground state expectation value, i.~e.\  $\vev{\Delta_{A}(x)}\ne 0$,
the color symmetry of the model (\ref{8}) is spontaneously broken
down.

It is more convenient to perform our investigations in terms of the
semi-bosonized Lagrangian (\ref{9}), since in this case we have a
common footing for obtaining both the thermodynamic potential and the
effective action of the model. Indeed, in the one fermion-loop
approximation, the effective action ${\cal S}_{\rm
{eff}}(\sigma,\pi_a,\Delta_{A},\Delta^{*}_{A'})$ of the model
(\ref{9}) is expressed by means of the path integral over quark
fields:
$$
\exp(i {\cal S}_{\rm {eff}}(\sigma,\pi_a,\Delta_{A},
\Delta^{*}_{A'}))=
  N'\int[d\bar q][dq]\exp\Bigl(i\int\tilde {\cal L}\,d^4 x\Bigr),
$$
where
\begin{eqnarray}
&&{\cal S}_{\rm {eff}}
(\sigma,\pi_a,\Delta_{A},\Delta^{*}_{A'})
=-\int d^4x\left [\frac{\sigma^2+\pi^2_a}{4G}+
\frac{\Delta_{A}\Delta^{*}_{A}}{4H}\right ]+
\tilde {\cal S}_{\rm {eff}},
\label{11}
\end{eqnarray}
and $N'$ is a normalization constant.
The quark contribution to the effective action, i.~e.\  the term
$\tilde {\cal S}_{\rm {eff}}$ in (\ref{11}), is given by:
\begin{equation}
\exp(i\tilde {\cal S}_{\rm {eff}})=N'\int [d\bar
q][dq]\exp\Bigl(\frac{i}{2}\int\Big [\bar
qD^+q+\bar q^CD^-q^C-\bar qK q^C-\bar
q^CK^{*}q\Big ]d^4 x\Bigr).
\label{12}
\end{equation}
In (\ref{12}) we have used the following notations
\begin{eqnarray}
&&D^+=i\gamma^\nu\partial_\nu- m+\hat\mu\gamma^0-\Sigma,~~~
D^-=i\gamma^\nu\partial_\nu- m-\hat\mu\gamma^0-\Sigma^t,
~~~ \Sigma=\sigma(x)+ i\gamma^5\pi_a(x)\tau_a,
\nonumber\\
&&~~~~~~~
\Sigma^t=\sigma(x)+ i\gamma^5\pi_a(x)\tau^t_a,~~~~~
K^*=i\Delta^{*}_{A}(x)\gamma^5\tau_2\lambda_{A},
\qquad
K=i\Delta_{A}(x)\gamma^5\tau_2\lambda_{A},
\label{13}
\end{eqnarray}
where $D^\pm$ are nontrivial operators in coordinate, spinor,
color and flavor spaces. \footnote{In order to bring the quark sector
of the Lagrangian (\ref{9}) to the expression, given in the square
brackets of (\ref{12}), we use the following well-known
relations:
$\partial_\nu^t=-\partial_\nu$, $C\gamma^\nu C^{-1}=-(\gamma^\nu)^t$,
$C\gamma^5C^{-1}=(\gamma^5)^t=\gamma^5$,
$\tau^2\vec\tau\tau^2=-(\vec\tau)^t$,
$\tau^2=\left (\begin{array}{cc}
0~, & -i\\
i~, &0
\end{array}\right )$.}
In the following, it is very convenient to use the Nambu--Gorkov
formalism, in which quarks are composed into a bispinor $\Psi$ such
that
\begin{equation}
\Psi=\left({q\atop q^C}\right),~~\Psi^t=(q^t,\bar q C^t);~~
\quad \bar\Psi=(\bar q,\bar q^C)=(\bar q,q^t C)=\Psi^t \left
(\begin{array}{cc}
0~~,&  C\\
C~~, &0
\end{array}\right )\equiv\Psi^t Y.
\label{14}
\end{equation}
Furthermore, by introducing the matrix-valued operator
\begin{equation}
Z=\left (\begin{array}{cc}
D^+, & -K\\
-K ^*, &D^-
\end{array}\right ),\label{15}
\end{equation}
one can rewrite the functional gaussian integral in (\ref{12}) in
terms of $\Psi$ and $Z$ and then evaluate it as follows
(clearly, in this case $[d\bar q][dq]=$ $[d q^C][dq]=$
$[d\Psi]$):
$$
\exp(i\tilde {\cal S}_{\rm {eff}})=  \int[d\Psi]\exp\left\{\frac
i2\int\bar\Psi Z\Psi d^4x\right\}=
  \int[d\Psi]\exp\left\{\frac i2\int\Psi^t(YZ)\Psi
  d^4x\right\}=\mbox {det}^{1/2}(YZ)=\mbox {det}^{1/2}(Z),
$$
where the last equality is valid due to the evident relation $\det
Y=1$. Then, using the general
formula $\det O=\exp {\rm Tr}\ln O$, one obtains the expression for
the effective action:
\begin{equation}
{\cal S}_{\rm
{eff}}(\sigma,\pi_a,\Delta_{A},\Delta^{*}_{A'})
=-\int d^4x\left[\frac{\sigma^2+\pi^2_a}{4G}+
\frac{\Delta_{A}\Delta^{*}_{A}}{4H}\right]-\frac i2{\rm
Tr}_{sfcx\scriptstyle NG}\ln Z.
\label{16}
\end{equation}
Besides of an evident trace over the two-dimensional Nambu --
Gorkov (NG) matrix, the trace in (\ref{16}) is
calculated  in spinor ($s$), flavor ($f$), color
($c$) and four-dimensional coordinate ($x$) spaces,
respectively.

Starting from (\ref{16}), one can define the quark contribution
$\Omega_q(\sigma,\pi_a, \Delta_{A}, \Delta^{*}_{A'})$ to the
thermodynamic potential (TDP) of the model (\ref{8}). In the
mean-field approximation one has
\begin{equation}
{\cal S}_{\rm {eff}}~\bigg
|_{~\sigma,\pi_a,\Delta_{A},\Delta^{*}_{A'}=\rm {const}}
=-\Omega_q(\sigma,\pi_a,\Delta_{A},\Delta^{*}_{A'})\int d^4x.
\label{17}
\end{equation}
The ground state expectation values (mean
values) of the fields:
$\vev{\sigma_a(x)}\equiv\sigma^o,~\vev{\pi_a(x)}
\equiv\pi_a^o,~\vev {\Delta_{A}(x)}\equiv\Delta^{o}_{A},~
\vev{\Delta^{*}_{A'}(x)}\equiv\Delta^{*o}_{A'}$, are
solutions of the gap equations for the TDP $\Omega_q$
(in our approach all ground state expectation values do not depend on
coordinates $x$):
\begin{eqnarray}
\frac{\partial\Omega_q}{\partial\pi_a}=0,~~~~~
\frac{\partial\Omega_q}{\partial\sigma}=0,~~~~~
\frac{\partial\Omega_q}{\partial\Delta_{A}}=0,~~~~~
\frac{\partial\Omega_q}{\partial\Delta^{*}_{A'}}=0.
\label{18}
\end{eqnarray}

Next, let us perform the following shift of bosonic fields in
(\ref{16}): $\sigma(x)\to\sigma(x)+\sigma^o$,
$\pi_a(x)\to\pi_a(x)+\pi_a^o$, $\Delta^{*}_{A}(x)
\to\Delta^{*}_{A}(x)+\Delta^{*o}_{A}$, $\Delta_{A}(x)
\to\Delta_{A}(x)+\Delta^{o}_{A}$. (Obviously, the new shifted
bosonic fields $\sigma (x),\pi_a (x),\Delta_{A}(x),
\Delta^{*}_{A}(x)$ now denote the small quantum fluctuations
around the mean values $\sigma^o,\pi^o_a,\Delta^{o}_{A},
\Delta^{*o}_{A}$ of mesons and diquarks rather than the original
fields (\ref{10})). In this case
\begin{equation}
Z=\left (\begin{array}{cc}
D^+_o~, & -K_o\\
 -K_o^*~, & D^-_o
\end{array}\right )-\left (\begin{array}{cc}
\Sigma~, & K\\
 K^*~, & \Sigma^t
\end{array}\right )\equiv S_0^{-1}-\left (\begin{array}{cc}
\Sigma~, & K\\
 K^*~, & \Sigma^t
\end{array}\right ),
\label{19}
\end{equation}
where $S_0$ is the quark propagator matrix in the Nambu -- Gorkov
representation (its matrix elements $S_{ij}$ are given in the
Appendix \ref{ApA}),
and
\[
(K_o, K^*_o, D^\pm_o,\Sigma_o,\Sigma^t_o)= (K, K^*, D^\pm,
\Sigma,\Sigma^t)~\bigg |_{~\sigma =\sigma^o, \pi_a =\pi_a^o, ...}
\]
Then, expanding the obtained expression into a Taylor-series up to
second order of small bosonic fluctuations, we have
\begin{equation}
 {\cal S}_{\rm
 {eff}}(\sigma,\pi_a,\Delta_{A},\Delta^{*}_{A'})=
 {\cal S}_{\rm {eff}}^{(0)} +
 {\cal S}_{\rm
 {eff}}^{(2)}(\sigma,\pi_a,\Delta_{A},\Delta^{*}_{A'})
 +\cdots,
  \label{20}
\end{equation}
where (due to the gap equations, the linear term
in meson and diquark fields is absent in (\ref{20}))
\begin{eqnarray}
 &&{\cal S}_{\rm {eff}}^{(0)}=-\int
 d^4x\left[\frac{\sigma^o\sigma^o+\pi^o_a\pi^o_a}{4G}+
\frac{\Delta^{o}_{A}\Delta^{*o}_{A}}{4H}\right]-\frac i2{\rm
Tr}_{scfx\scriptstyle NG}\ln
 \left (S_0^{-1}\right )\nonumber\\
&&~~~~~~~~~~~~~~~~~~~~~~~~~~~~~
\equiv-\Omega_q(\sigma^o,\pi^o_a,\Delta^{o}_{A},
 \Delta^{*o}_{A'})\int d^4x,
  \label{21}\\
 {\cal S}^{(2)}_{\rm
 {eff}}(\sigma,\pi_a,\Delta_{A},\Delta^{*}_{A'})
 \!\!\!\!&&\!\!=
 -\int d^4x\left[\frac{\sigma^2+\pi^2_a}{4G}+
\frac{\Delta_{A}\Delta^{*}_{A}}{4H}\right]\nonumber\\
&&~~~~~~~~~~~~~~~~~~~~~~+\frac i4{\rm
Tr}_{scfx\scriptstyle NG}
\left\{S_0\left (\begin{array}{cc}
\Sigma~, & K\\
 K^*~, & \Sigma^t
\end{array}\right )S_0\left (\begin{array}{cc}
\Sigma~, & K\\
 K^*~, & \Sigma^t
\end{array}\right )\right\}.
  \label{22}
\end{eqnarray}
In the following
we will study the spectrum of meson/diquark excitations
in the color superconducting phase of the NJL model on the basis of
the effective action ${\cal S}_{\rm {eff}}^{(2)}$. The effective
action (\ref{22}) can be presented in the explicit form:
\begin{eqnarray}
{\cal S}^{(2)}_{\rm {eff}}={\cal S}^{(2)}_{\rm mesons}+{\cal
S}^{(2)}_{\rm diquarks}+{\cal S}^{(2)}_{\rm mixed},
\label{23}
\end{eqnarray}
where
\begin{eqnarray}
\label{24}
 {\cal S}^{(2)}_{\rm mesons} \!\!\!\!&&\!\!=
 -\int d^4x\frac{\sigma^2+\pi^2_a}{4G}+
\frac i4{\rm Tr}_{scfx}
\left\{S_{11}\Sigma S_{11}\Sigma +2S_{12}\Sigma^tS_{21}\Sigma +
S_{22}\Sigma^t S_{22}\Sigma^t\right\},\\
\label{25}
 {\cal S}^{(2)}_{\rm diquarks} \!\!\!\!&&\!\!=
 -\int d^4x\frac{\Delta_{A}\Delta^{*}_{A}}{4H}
+\frac i4{\rm Tr}_{scfx}
\left\{S_{12}K^*S_{12}K^* +2S_{11}KS_{22}K^* +
S_{21}K S_{21}K\right\},\\
 {\cal S}^{(2)}_{\rm mixed} \!\!\!\!&&\!\!=
\frac i2{\rm Tr}_{scfx}
\left\{S_{11}\Sigma S_{12}K^* +S_{21}\Sigma S_{11}K +
S_{12}\Sigma^t S_{22}K^*+S_{21}KS_{22}\Sigma^t\right\},
 \label{26}
\end{eqnarray}
and $S_{ij}$ are the matrix elements of the quark propagator matrix
$S_0$ defined in (\ref{19}) (see also Appendix \ref{ApA}). Moreover,
some necessary explanations concerning the trace-operation over
coordinate space in the expressions (\ref{24})-(\ref{26}) are given
in Appendix \ref{ApB} (see (\ref{B4})). It follows from these
formulae that the effective action (\ref{24}) is a functional of the
meson fields $\sigma(x),\pi_i(x)$ only, the effective action
(\ref{25}) is composed from diquark fields only, and the mixing
between mesons and diquarks might occur because of the effective
action (\ref{26}).

\section{Gap equations and neutrality conditions}
\label{III}

Earlier (see, e.g., the papers \cite{sh2,zhuang,bs}), it was shown
that for electrically and color neutral cold dense matter,
described in the framework of the model (\ref{3}), only two phases
are allowed to exist. In the first one, the symmetric phase that
is usually called the normal quark matter phase, only the mean
value of the $\sigma$-field, $\vev{\sigma(x)}\equiv\sigma^o$, is
nonzero. In the second one which is just the 2SC phase of dense
matter the mean value of the diquark field $\Delta_{2}(x)$ is
nonzero in addition, i.~e.\  $\vev{\Delta_{2}(x)}
\equiv\Delta^{o}_{2}\ne 0$. Hence, without loss of generality, it
is convenient to deal with TDP $\Omega_q(\sigma^o,\pi^o_a,
\Delta^{o}_{A},\Delta^{*o}_{A'})$ (note, $A,A'=2,5,7$), in which
all arguments, except $\sigma^o\equiv M-m$ ($m$ is a bare quark
mass, whereas the parameter $M$ is usually called constituent or
dynamical quark mass) and $\Delta^{o}_{2}\equiv\Delta$, are
identically equal to zero. In this case the calculation of the
total TDP $\Omega$ of the system (\ref{3}) (recall,
$\Omega=\Omega_e+ \Omega_q$) is significantly simplified
\cite{sh2,zhuang}, and we have
\begin{eqnarray}
&&\Omega(\mu_B,\mu_Q,\mu_8;M,\Delta)=
-\frac{\mu_Q^4}{12\pi^2}+\frac{(M-m)^2}{4G}+\frac{|\Delta|^2}{4H}
\nonumber\\
&&-2\sum_{\pm}\int\frac{d^3q}{(2\pi)^3}\Big\{|E_{\Delta}^\pm+
\delta\mu|+|E_{\Delta}^\pm-\delta\mu|\Big\}-
\sum_{\pm}\int\frac{d^3q}{(2\pi)^3}\Big\{
|E\pm\mu_{ub}|+|E\pm\mu_{db}|\Big\},
\label{27}
\end{eqnarray}
where $E_\Delta^\pm=\sqrt{(E^\pm)^2+|\Delta|^2}$,
$E^\pm=E\pm\bar\mu$, $E=\sqrt{\vec q^2+M^2}$,
$\bar\mu=(\mu_{ur}+\mu_{dg})/2=$$(\mu_{ug}+\mu_{dr})/2=$$\tilde\mu+
\mu_8=$$\mu_B/3+\mu_Q/6+\mu_8$,
$\mu_{ub}=\mu_B/3+2\mu_Q/3-2\mu_8$,
$\mu_{db}=\mu_B/3-\mu_Q/3-2\mu_8$, $\delta\mu=\mu_Q/2$ (see also
the notations, used in (\ref{7})). Note that, apart from the order
parameters $M$ and $\Delta$, the TDP $\Omega$ really depends on
the chemical potentials, which is indicated in (\ref{27}) in an
explicit form. Since the integrals in the right hand side of
(\ref{27}) are ultraviolet divergent, we regularize them as well
as the other three-dimensional divergent integrals below by
implementing a cutoff in the integration regions, $|\vec
q|<\Lambda$. Starting from (\ref{27}), one can find the gap
equations $\partial\Omega /\partial\Delta^* =0$ and
$\partial\Omega /\partial M =0$, which supply us with the values
of $M,\Delta$ in the ground state of the system:
\begin{eqnarray}
&&\frac{\partial\Omega}{\partial\Delta^*}\equiv\frac{\Delta}{4H}-
2\Delta\int\frac{d^3q}{(2\pi)^3}
\left[\frac{\theta(E_\Delta^+-|\delta\mu|)}{E_\Delta^+}+
\frac{\theta(E_\Delta^--|\delta\mu|)}{E_\Delta^-}\right]=0,
\label{28}\\
&&\frac{\partial\Omega}{\partial M}\equiv\frac{M-m}{2G}-
2M\int\frac{d^3q}{(2\pi)^3E}
\Big [\theta(E-|\mu_{ub}|)+
\theta(E-|\mu_{db}|)\Big ]\nonumber\\
&&~~~~~~~~~~~~~~~~~~~~-4M\int\frac{d^3p}{(2\pi)^3E}\left
[\frac{\theta(E_\Delta^+
-|\delta\mu|)E^+}{E_\Delta^+}
+\frac{\theta(E_\Delta^--|\delta\mu|)E^-}{E_\Delta^-}\right ]=0.
\label{29}
\end{eqnarray}
Next, let us impose the local color- as well as electric charge
neutrality requirements on the ground state of the model (\ref{3}).
It means that the quantities $\mu_8$ and $\mu_Q$ take such values
that the densities of the 8-th color charge $N_8$ and electric charge
$N_Q$ are equal to zero in the ground state for arbitrary fixed
values of other model parameters, i.~e.\  $\vev{N_8}=-\partial\Omega
/\partial\mu_8\equiv 0$, $\vev{N_Q}= -\partial\Omega
/\partial\mu_Q\equiv 0$. These neutrality constraints look like:
\begin{eqnarray}
&&\vev{N_8}\equiv 4\int\frac{d^3q}{(2\pi)^3}
\Big [\frac{\theta(E_\Delta^+
-|\delta\mu|)E^+}{E_\Delta^+}
-\frac{\theta(E_\Delta^--|\delta\mu|)E^-}{E_\Delta^-}
\Big ]
\nonumber\\&&~~~~~~-4\int\frac{d^3p}{(2\pi)^3}
\Big [sign(\mu_{ub})\theta(|\mu_{ub}|-E)+
sign(\mu_{db})\theta(|\mu_{db}|-E)\Big ]=0,
\label{30}\\
&&\vev{N_Q}\equiv\frac{\mu_Q^3}{3\pi^2}+\frac{\vev{N_8}}{6}
+2\int\frac{d^3q}{(2\pi)^3}
\Big [sign(\mu_{ub})\theta(|\mu_{ub}|-E)
\nonumber\\ &&
~~~~~~~~~~~~~~+sign(\delta\mu)\theta(|\delta\mu|-E_\Delta^+)+
sign(\delta\mu)\theta(|\delta\mu|-E_\Delta^-)\Big ]=0,
\label{31}
\end{eqnarray}
In the following we suppose, for simplicity, that $\Delta$ is a real
quantity. Of course, the solution of the common system of gap
equations (\ref{28})-(\ref{29}) and neutrality relations 
(\ref{30})-(\ref{31}) is possible only numerically. In all numerical
calculations of the present paper we use the following parameter set
\begin{eqnarray}
G=5.86~{\rm GeV}^{-2},~~~~~\Lambda=618~{\rm MeV},~~~~~m=5.67~ {\rm
MeV}
\label{32}
\end{eqnarray}
that leads in the framework of the NJL model to the well-known
vacuum phenomenological values of the pion weak-decay constant
$F_\pi=92.4$~MeV, pion mass $M_\pi=140$~MeV, and chiral quark
condensate $\vev{\bar qq}=-(245~\mbox{MeV})^3$. Moreover, we use
two different values for the coupling constant $H$ in the diquark
channel, $H=3G/4$ and $H=G$, for which two qualitatively different
2SC phases are realized in the model (see below). The numerical
analysis shows that for the parameter set (\ref{32}) and both
relations $H=3G/4$ and $H=G$ the system of equations
(\ref{28})-(\ref{31}) has only two solutions. As was already
discussed after (\ref{7}), the first one (with $M\ne 0$,
$\Delta=0$, $\mu_8=0$ and $\mu_Q=0$) corresponds to the
SU(3)$_c\times$SU(2)$_I$ symmetry of the model (normal phase),
whereas the second one (with $M\ne 0$, $\Delta\ne 0$, $\mu_8\ne 0$
and $\mu_Q\ne 0$) corresponds to the 2SC phase. As usual,
solutions of these equations give  local extrema of the
thermodynamic potential $\Omega(\mu_B,\mu_Q,\mu_8;M,\Delta)$
(\ref{27}). Clearly, one should also check in which of them the
TDP takes the least value (for each fixed value of $\mu_B$).
Just this solution of the system of equations (\ref{28})-(\ref{31})
corresponds to the genuine neutral ground state of the model. In
particular, for each fixed $\mu_B$ it supplies us with mean values
(gaps) $M,\Delta$ (see Figs 1,2) as well as with values $\mu_8$ and
$\mu_Q$ (see Figs 3,4), at which the 
ground state has zero charges.\footnote{\label{f4} In the literature,
the zero value of the current quark mass $m$ is of frequent use in
the 2SC investigation. In this case the constituent quark mass $M$ is
identically equal to zero in the 2SC phase of neutral matter (see,
e.g., in \cite{sh2,he}).}
\begin{figure}
 \includegraphics[width=0.45\textwidth]{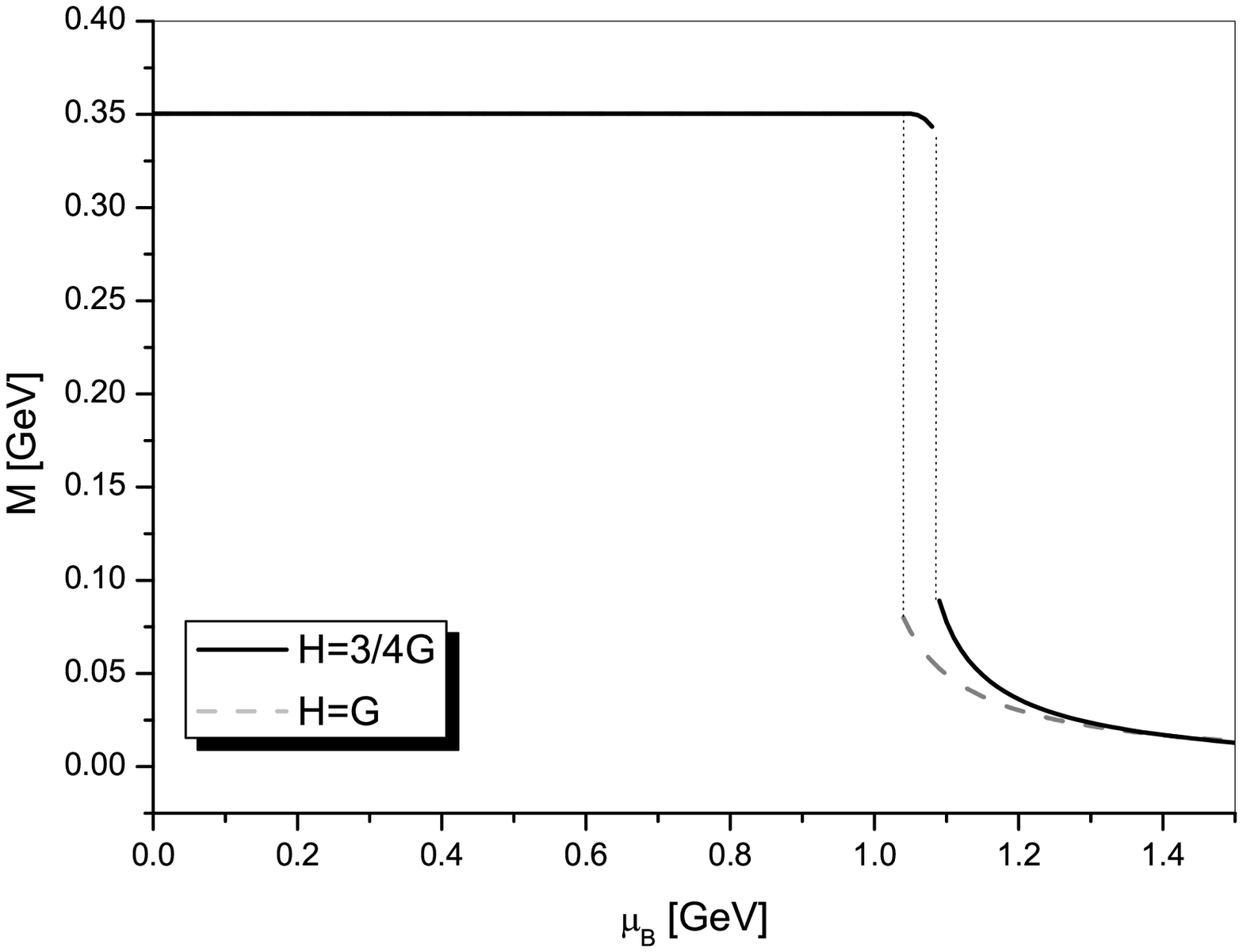}
 \hfill
 \includegraphics[width=0.45\textwidth]{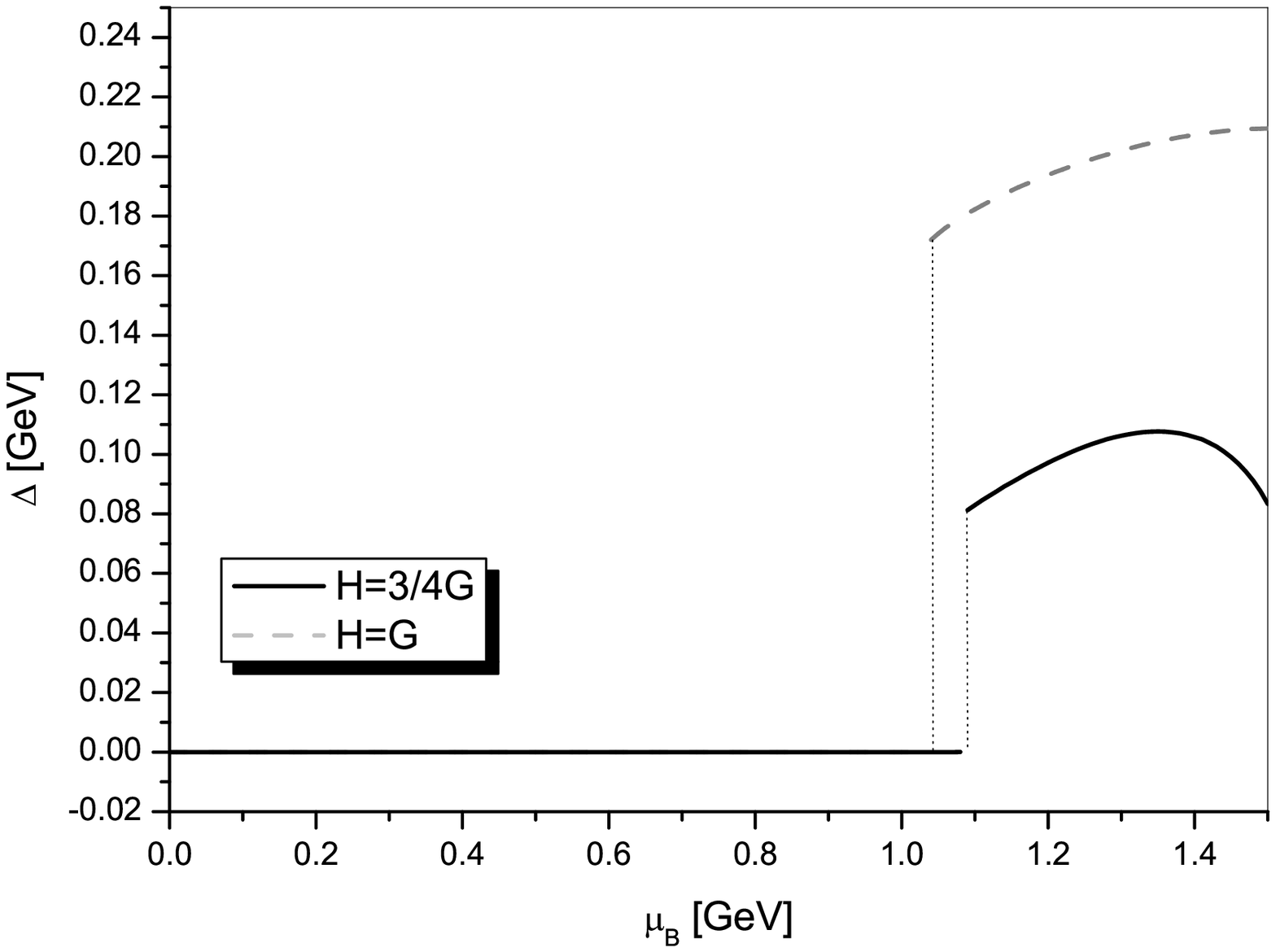}\\
\parbox[t]{0.45\textwidth}{
 \caption{The behaviour of $M$ vs $\mu_B$ in the neutral matter.}
 \label{fig:1}
 }
\hfill
\parbox[t]{0.45\textwidth}{
\caption{The behaviour of $\Delta$ vs $\mu_B$ in the neutral
matter.} \label{fig:2} }
\end{figure}

It is clear from Figs 1,2 that there is a critical value of the
baryon chemical potential $\mu_B^c$ (for the case $H=3G/4$ one has
$\mu_B^c=1.08$ GeV, whereas for the case $H=G$ one gets
$\mu_B^c=1.04$ GeV \footnote{The tendency, the greater $H$ the
smaller $\mu_B^c$, is indeed supported by earlier investigations
of the 2SC phenomenon in the framework of NJL models
\cite{hzc,ek}. In particular, it was shown that at sufficiently
large values of the coupling constant $H$ the 2SC phase may be
realized in the model even at zero baryon chemical potential
$\mu_B$ \cite{ek}.}) such that at $\mu_B<\mu_B^c$ the normal
SU(3)$_c\times$SU(2)$_I$-symmetric phase of the model occurs.
However, at $\mu_B>\mu_B^c$ the 2SC phase is realized. Since in
this case $\mu_Q\ne 0$ (see Fig. 4), the isospin
SU(2)$_I$-symmetry of the normal phase is broken in the critical
point rather by hand than dynamically, down to the U(1)$_{I_3}$
group. Furthermore, in the 2SC phase we have $\vev{\Delta_{2}(x)}
=\Delta\ne 0$, $\vev{\Delta_{5,7}(x)}\equiv 0$. So in the critical
point $\mu_B^c$ the color SU(3)$_c$-symmetry is broken down to the
SU(2)$_c$ subgroup. Since at $\mu_B>\mu_B^c$ the $\mu_8$-term of
the Lagrangian (\ref{3}) is nonzero (see Fig. 3), the initial
color symmetry of the model is also broken by hand down to the
SU(2)$_c\times$U(1)$_{\lambda_8}$ subgroup, which is further
broken dynamically (spontaneously) down to SU(2)$_c$. Hence, one
may expect the appearance of only one Nambu -- Goldstone boson in
the mass spectrum of the 2SC phase (see the discussion below).
Finally, remark that in the critical point $\mu_B^c$ the
transition between these two phases is of the first-order which is
characterized by a discontinuity in the behavior of $M$ and
$\Delta$ vs. $\mu_B$ (see Figs 1,2).

Up to now we have discussed the properties of the 2SC phase in
neutral and $\beta$-equilibrated matter which are common for the
two particular cases $H=3G/4$ and $H=G$. However, there exist, at
least two, qualitative distinctions between these neutral 2SC
phases. (Both of them are based on the fact that in the neutral
2SC matter the relation $\Delta<|\delta\mu|$ is valid for the case
of $H=3G/4$, whereas at $H=G$ the opposite one
$\Delta>|\delta\mu|$ is true.) The first one lies in the diquark
mass spectrum and will be discussed below. Just now we would like
to present the second one which is provided by the quasiparticle
dispersion relations, i.~e.\  the momentum dependence of energy.
In condensed matter physics quasiparticles are simply the
one-fermion excitations of the ground state. In our case the
quasiparticle spectrum of the 2SC matter is defined by
singularities of the quark propagator $S_0$ (see Appendix \ref{ApA}).
Clearly, there are twelve (6 quark and 6 antiquark) quasiparticles
in the 2SC matter. Four of them, blue quasiparticles, have the
energies $E\pm\mu_{ub}$ and $E\pm\mu_{db}$. The energy spectrum of
the other eight, red and green quasiparticles, consists of four
values $E_\Delta^\pm\pm |\delta\mu|$, each is doubly degenerate.
Evidently, for both relations between coupling constants $H$ and
$G$ there are momentum values at which the energies of two blue
quasiparticles must turn into zero, i.~e.\  there are no energy
costs to create these quasiparticles (their energies are
$E-\mu_{ub}$ and $E-\mu_{db}$). Due to this reason, these
excitations are called gapless ones. Similarly, at $H=3G/4$ there
are in addition also two gapless red and green fermionic
excitations (with energies $E_\Delta^--|\delta\mu|$) of neutral
2SC matter, since in this case the relation $\Delta<|\delta\mu|$
is true. However, in the neutral 2SC matter of the case $H=G$
there are no additional gapless red and green quasiparticles.
Hence, at $H=3G/4$ the quasiparticle spectrum of the neutral 2SC
matter consists of four gapless excitations (it is a so-called
gapless color superconductivity). In contrast, at $H=G$ there are
only two gapless quasiparticles in the neutral 2SC matter, and
color superconductivity is called as gapped one. Thus, our results
corroborate the conclusion, made, e.g., in \cite{sh2,abuki}, that
the gapless 2SC may exist for a rather narrow interval of the
coupling constant $H$ values.

In the next sections we will calculate the inverse two-point
(unnormalized) correlators of meson and diquark fluctuations over
the ground state of neutral 2SC matter in the one-loop
(mean-field) approximation and find their masses in the two cases,
$H=3G/4$ and $H=G$.
\begin{figure}
\includegraphics[width=0.45\textwidth]{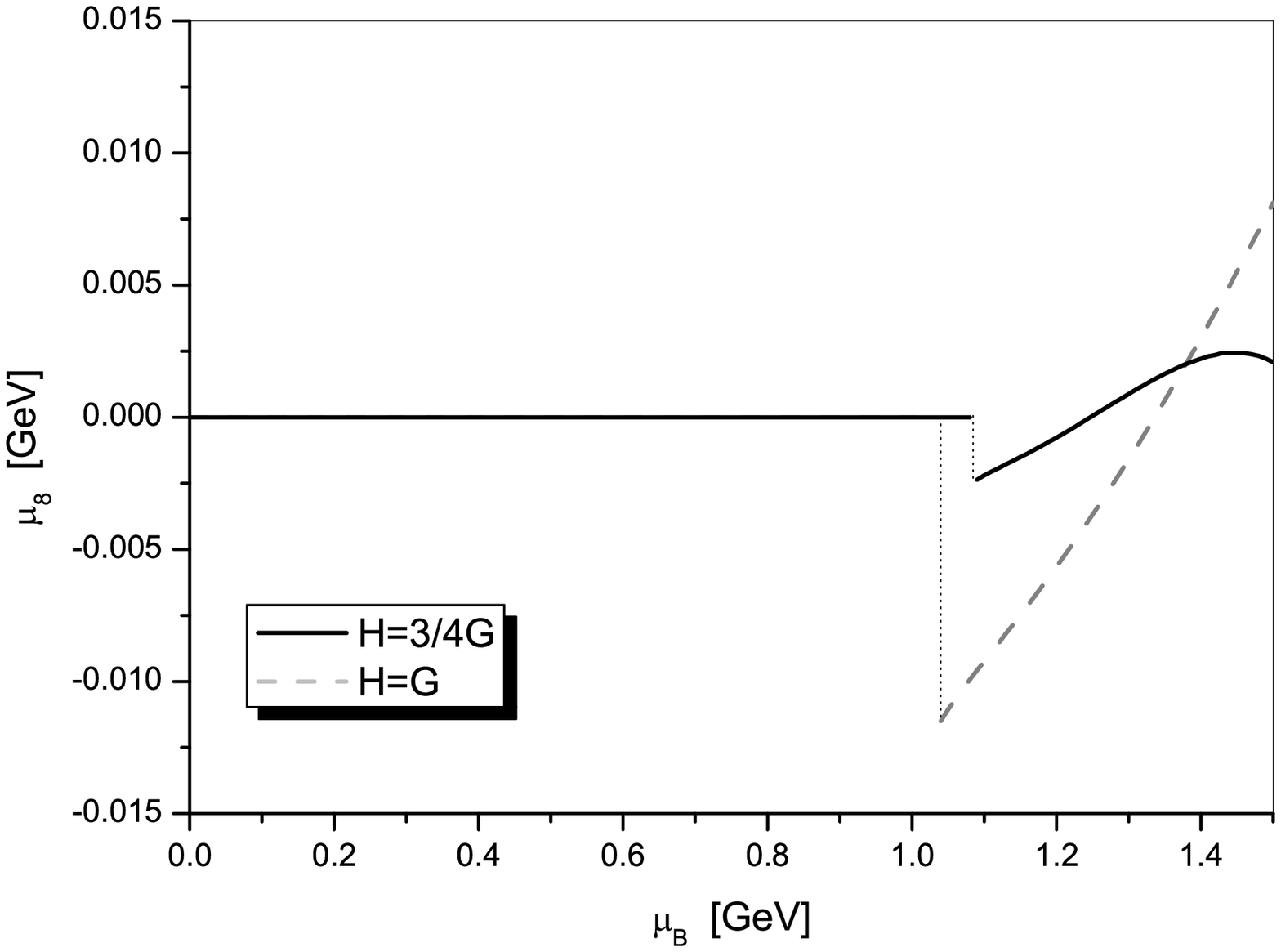}
\hfill
\includegraphics[width=0.45\textwidth]{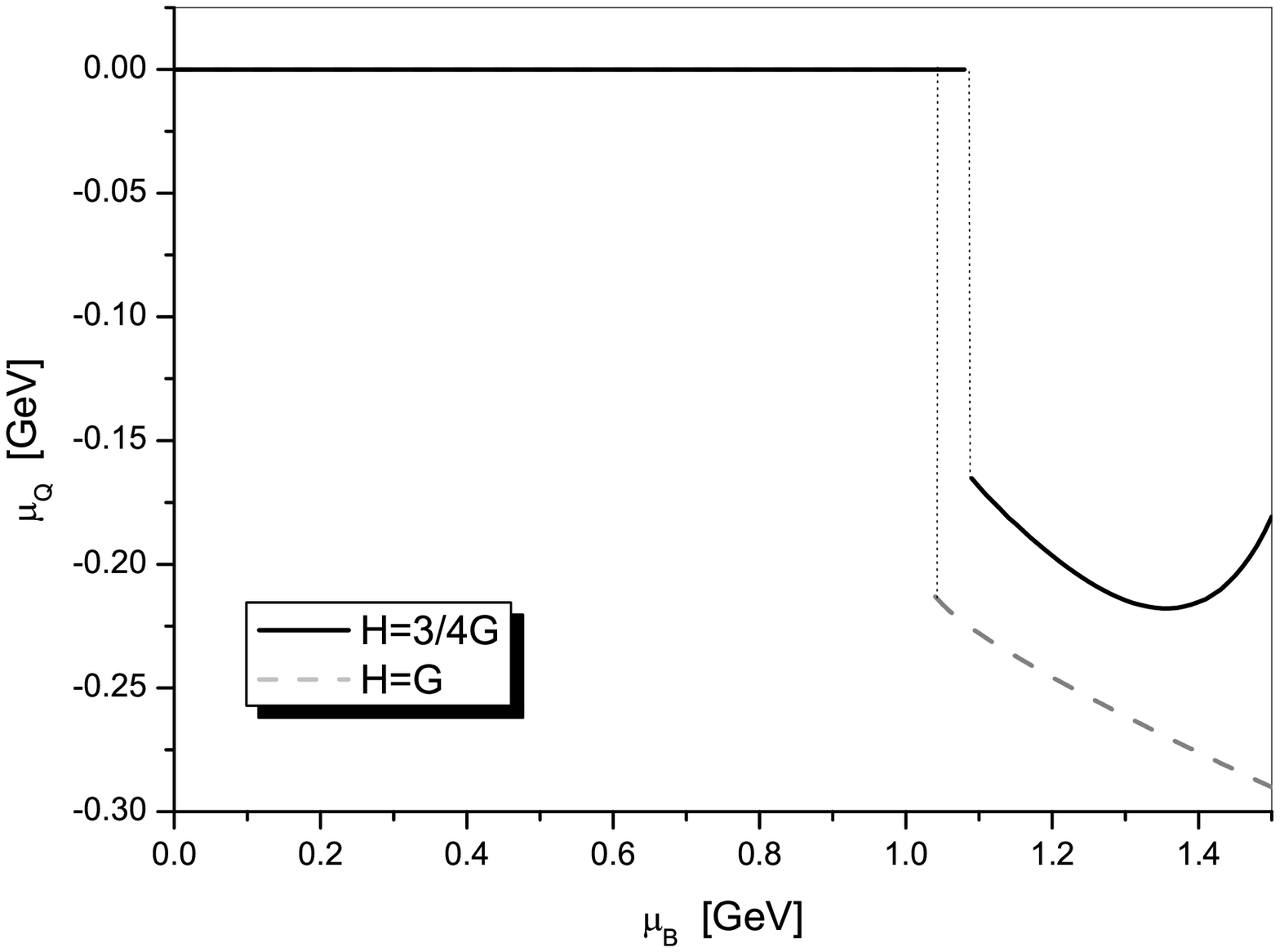}
\parbox[t]{0.45\textwidth}{
\caption{The behaviour of $\mu_8$ vs $\mu_B$ in the neutral
matter.}
 \label{fig:3}
 }\hfill
\parbox[t]{0.45\textwidth}{
\caption{The behaviour of $\mu_Q$ vs $\mu_B$ in the neutral
matter.} \label{fig:4} }
\end{figure}

\section{Meson masses}

After a more detailed study of the ${\cal S}^{(2)}_{\rm mixed}$ -part
(\ref{26}) of the effective action, it turns out that it is composed
from $\sigma (x)$, $\Delta_{2}(x)$ and $\Delta^{*}_{2}(x)$ fields
only, i.~e.\  the $\pi$-mesons are not mixed with diquarks (this
property is justified by the parity conservation both in the normal-
and 2SC phases of our model). So to find the $\pi$-meson masses it is
enough to deal with the effective action ${\cal S}^{(2)}_{\rm
mesons}$ (\ref{24}), which is a generating functional of the
one-particle irreducible (1PI) Green functions of $\sigma$- and
$\pi$-meson fields. In this case, instead of $\pi_i(x)$-fields
(\ref{10}), we will use the new fields $\pi^0(x)\equiv\pi_3(x)$,
$\pi^\pm(x)\equiv (\pi_1(x)\pm\pi_2(x))/\sqrt{2}$, so that the
quantities $\Sigma,\Sigma^t$ from (\ref{13}) look like:
\begin{eqnarray}
\Sigma = \sigma(x)+i\gamma^5[\pi^0(x)\tau_3+\pi^-(x)
\tau_++\pi^+(x)\tau_-],~~~
\Sigma^t = \sigma(x)+i\gamma^5[\pi^0(x)\tau_3+\pi^-(x)
\tau_-+\pi^+(x)\tau_+],
 \label{33}
\end{eqnarray}
where $\tau_\pm\equiv (\tau_1\pm\tau_2)/\sqrt{2}$.
Then, the 1PI Green functions of $\sigma$- and
$\pi^\pm,\pi^0$-meson fields can be generated through the relation:
\begin{eqnarray}
&& \Gamma_{XY}(x-y)=-\frac{\delta^2{\cal S}^{(2)}_{\rm
mesons}}{\delta Y(y)\delta X(x)},
 \label{34}
\end{eqnarray}
where $X,Y=\sigma,\pi^\pm,\pi^0$ (to take the variational derivatives
in (\ref{34}), it is very instructive to refer to the relations
(\ref{B4})-(\ref{B5})). In momentum space the zeros of the Fourier
transformations of these functions are connected with meson masses.

\subsection{The $\pi^\pm$-meson masses}
\label{pipm}

Using the relation (\ref{34}), it is possible to define the 1PI Green
functions of the $\pi^\pm$-fields. In particular,
\begin{eqnarray}
&& \Gamma_{\pi^+\pi^-}(x-y)=\frac{\delta(z)}{2G}+2i{\rm Tr}_{sc}
\left[2A_{11}(z)\gamma^5B_{11}(-z)\gamma^5+2B_{22}(z)\gamma^5A_{22}
(-z)\gamma^5\right.\nonumber\\&&~~~~~~~~~~~~~~~~\left.
+A_{12}(z)\gamma^5A_{21}(-z)\gamma^5+B_{21}(z)\gamma^5
B_{12}(-z)\gamma^5\right]
\label{35}
\end{eqnarray}
(in this expression all traces over flavor indices are calculated,
and $z=x-y$), where operators $A_{ij}(z)$ and $B_{kl}(z)$ are defined
in Appendix \ref{ApA}. Moreover, $\Gamma_{\pi^-\pi^{+}}(x-y)=
\Gamma_{\pi^+\pi^{-}}(y-x)$ and $\Gamma_{\pi^+\pi^{+}}(x-y)=
\Gamma_{\pi^-\pi^{-}}(x-y)=0$.
Due to the traces, containing an odd number of $\gamma^5$, all the
mixed Green functions of the form
$\Gamma_{\sigma\pi^{0,\pm}}(x-y)=-\frac{\delta^2{\cal S}^{(2)}_{\rm
mesons}}{\delta\pi^{0,\pm}(y)\delta\sigma(x)}$ are zero. Moreover,
since the traces containing an odd number of $\tau_\pm$-matrices
equal zero, all the 1PI Green functions of the form
$\Gamma_{\pi^0\pi^{\pm}}(x-y)$ are also zero. Hence, in the framework
of our model there is no any mixing between $\pi^\pm$-fields on the
one hand, and $\sigma,\pi^0$-fields on the other. After taking the
trace over
color indices, the Fourier transform of (\ref{35}) has the following
form:
\begin{eqnarray}
&&\overline{\Gamma_{\pi^+\pi^-}}(p)=\frac{1}{2G}+4i{\rm Tr}_{s}
\int\frac{d^4q}{(2\pi)^4}\left[2\overline{
a^{(12)}_{11}}(p+q)\gamma^5\overline{b^{(12)}_{11}}(q)\gamma^5+
\overline{a^{(3)}_{11}}(p+q)\gamma^5\overline{b^{(3)}_{11}}(q)
\gamma^5+\right.\nonumber\\&&~~\left.
2\overline{b^{(12)}_{22}}(p+q)\gamma^5\overline{a^{(12)}_{22}}(q)
\gamma^5+
\overline{b^{(3)}_{22}}(p+q)
\gamma^5\overline{a^{(3)}_{22}}(q)\gamma^5+
\overline{a_{12}}(p+q)\gamma^5\overline{a_{21}}(q)\gamma^5+
\overline{b_{21}}(p+q)\gamma^5\overline{b_{12}}(q)
\gamma^5\right],
\label{36}
\end{eqnarray}
where the quantities $\overline{a^{(12)}_{11}}(q)$,
$\overline{b^{(12)}_{11}}(q)$ etc are the corresponding Fourier
transformations of $a^{(12)}_{11}(z)$,
$b^{(12)}_{11}(z)$ etc, presented in Appendix \ref{ApA}. Clearly,
$\overline{\Gamma_{\pi^-\pi^+}}(p)=$$\overline
{\Gamma_{\pi^+\pi^-}}(-p)$.
The zeros of these functions determine the $\pi^\pm$-meson
dispersion laws, i.~e.\  the relations between their energy
and three-momenta. In the present paper, we are mainly interested
in the investigation of the modification of meson and diquark masses
in dense and cold color- and electrically neutral matter. Since in
this case a particle mass is defined as the value of its energy in
the rest frame, $\vec p=0$ (see, e.g., \cite{eky1,eky2,ruivo}), we
put $p=(p_0,0,0,0)$ in the following. As a result, the
calculation of 1PI Green functions is significantly simplified.
Indeed, in the rest frame one can easily perform all the trace
calculations over spinor indices in (\ref{36}) (see the auxiliary
relations (\ref{A10})) and gets
\begin{eqnarray}
&&\overline{\Gamma_{\pi^+\pi^-}}(p_0)=\frac{1}{2G}
-16i\int\frac{d^4q}{(2\pi)^4}
\frac{(q_0-\delta\mu)(p_0+q_0+\delta\mu)-E^+E^--|\Delta|^2}
{[(q_0-\delta\mu)^2-(E_\Delta^-)^2][(p_0+q_0+\delta\mu)^2-
(E_\Delta^+)^2]}\nonumber\\
&&-4i\int\frac{d^4q}{(2\pi)^4}\left[
\frac{1}{(p_0+q_0+\mu_{ub}+E)(q_0+\mu_{db}-E)}+\frac{1}{(p_0+q_0+
\mu_{ub}-E)(q_0+\mu_{db}+E)}\right],
\label{37}
\end{eqnarray}
where we have used the same notations as in (\ref{27}). Note
also that in (\ref{37}) $q_0$ is a shorthand notation for
$q_0+i\varepsilon\cdot {\rm sign}(q_0)$ and $(p_0+q_0)$ is a
shorthand notation for $(p_0+q_0)+i\varepsilon\cdot {\rm
sign}(p_0+q_0)$, where $\varepsilon\to 0_+$ (see also the remark
after (\ref{A9}) and \cite{chodos}). The
$q_0$-integration in (\ref{37}) is performed along the real axis in
the complex $q_0$-plane. We will close this contour by an infinite
arc in the upper half of the complex $q_0$-plane. Inside the obtained
closed contour the integrand of the first integral in (\ref{37}) has
four simple poles which are located in the following points:
\begin{eqnarray*}
(q_0)_1=\delta\mu-E_\Delta^-+i\varepsilon\cdot\theta(E_\Delta^-
-\delta\mu),~~~~&&~~~~(q_0)_2=\delta\mu+E_\Delta^-+i\varepsilon\cdot
\theta(-E_\Delta^--\delta\mu),\nonumber\\
(q_0)_3=E_\Delta^+-\delta\mu-p_0+i\varepsilon\cdot\theta(\delta\mu-
E_\Delta^+),~~~~&&~~~~(q_0)_4=-E_\Delta^+-\delta\mu-p_0+
i\varepsilon\cdot\theta(\delta\mu+ E_\Delta^+),
\end{eqnarray*}
whereas the integrand in the second line of (\ref{37}) has the
following four poles in the upper half of the $q_0$-plane:
\begin{eqnarray*}
(\breve
q_0)_1=E-\mu_{db}+i\varepsilon\cdot\theta(\mu_{db}-E),~~~~&&~~~~
(\breve q_0)_2=-\mu_{db}-E+i\varepsilon\cdot
\theta(\mu_{db}+E),\nonumber\\
(\breve
q_0)_3=-E-\mu_{ub}-p_0+i\varepsilon\cdot\theta(E+\mu_{ub}),~~~~&&~~~~
(\breve q_0)_4=E-\mu_{ub}-p_0+
i\varepsilon\cdot\theta(\mu_{ub}-E).
\end{eqnarray*}
Summing the residues of the integrand function in these poles, we can
perform the $q_0$-integration in (\ref{37}) and obtain:
\begin{eqnarray}
&&~~~~~~~~\overline{\Gamma_{\pi^+\pi^-}}(p_0)=\frac{1}{2G}
+8\int\frac{d^3q}{(2\pi)^3}\left[\frac{\theta(-\delta\mu-
E_\Delta^-)}{E_\Delta^-}\cdot
\frac{(p_0+E_\Delta^-+2\delta\mu)E_\Delta^--E^+E^--|\Delta|^2}
{(p_0+E_\Delta^-+2\delta\mu)^2-(E_\Delta^+)^2}+\right.\nonumber\\
&&\frac{\theta(E_\Delta^--\delta\mu)}{E_\Delta^-}\cdot
\frac{(p_0-E_\Delta^-+2\delta\mu)E_\Delta^-+E^+E^-+|\Delta|^2}
{(p_0-E_\Delta^-+2\delta\mu)^2-(E_\Delta^+)^2}-
\frac{\theta(\delta\mu-E_\Delta^+)}{E_\Delta^+}\cdot
\frac{(p_0-E_\Delta^++2\delta\mu)E_\Delta^++E^+E^-+|\Delta|^2}
{(p_0-E_\Delta^++2\delta\mu)^2-(E_\Delta^-)^2}\nonumber\\
&&~~~~~~\left.-\frac{\theta(\delta\mu+E_\Delta^+)}{E_\Delta^+}\cdot
\frac{(p_0+E_\Delta^++2\delta\mu)E_\Delta^+-E^+E^--|\Delta|^2}
{(p_0+E_\Delta^++2\delta\mu)^2-(E_\Delta^-)^2}\right]+
4\int\frac{d^3q}{(2\pi)^3}\left[\frac{\theta(\mu_{db}-
E)}{p_0+2\delta\mu+2E}\right.\nonumber\\
&&~~~~~~~~~~~~\left.-\frac{\theta(\mu_{ub}+E)}{p_0+2\delta\mu+2E}
+\frac{\theta(E+\mu_{db})}{p_0+2\delta\mu-2E}-
\frac{\theta(\mu_{ub}-E)}{p_0+2\delta\mu-2E}
\right].
\label{38}
\end{eqnarray}
Clearly, the matrix element $\overline{\Gamma_{\pi^+\pi^-}}(p_0)$
depends effectively on the variable $z=(p_0+2\delta\mu)$. It follows
from our numerical analysis that for both values of the coupling
constant $H$ ($H=G$ and $H=3G/4$) the expression (\ref{38}) has only
two zeros, $z_1(\mu_B)$ and $z_2(\mu_B)$. Hence, for each fixed value
of $\mu_B>\mu_B^c$, i.~e.\  in the 2SC phase, we have
$\overline{\Gamma_{\pi^+\pi^-}}(p_0)\sim (p_0+2\delta\mu-z_1)
(p_0+2\delta\mu-z_2)$. Since
$\overline{\Gamma_{\pi^-\pi^+}}(p_0)=\overline{\Gamma_{\pi^+\pi^-}}
(-p_0)$, the determinant of the inverse propagator matrix of the
$\pi^\pm$-mesons has the following form:
\begin{eqnarray}
\overline{\Gamma_{\pi^+\pi^-}}(p_0)\cdot\overline{\Gamma_{\pi^-\pi^+}
} (p_0)\sim (p_0^2-(2\delta\mu-z_1)^2)(p_0^2-(2\delta\mu-z_2)^2).
\label{39}
\end{eqnarray}
Evidently, in the $p_0^2$-plane it turns into zero in two points,
which are the mass squared of the $\pi^\pm$-mesons. Hence, in the 2SC
neutral dense matter the $\pi^\pm$-mesons have different masses:
\begin{eqnarray}
M_{\pi^+}^2=(2\delta\mu-z_1)^2,~~~~~M_{\pi^-}^2=(2\delta\mu-z_2)^2.
\label{40}
\end{eqnarray}
The behaviour of $M_{\pi^+}$ and $M_{\pi^-}$ vs $\mu_B$ in the color
superconducting and neutral matter is depicted in Figs 5,6 for
the cases $H=3G/4$ and $H=G$, respectively.
\begin{figure}
\includegraphics[width=0.45\textwidth]{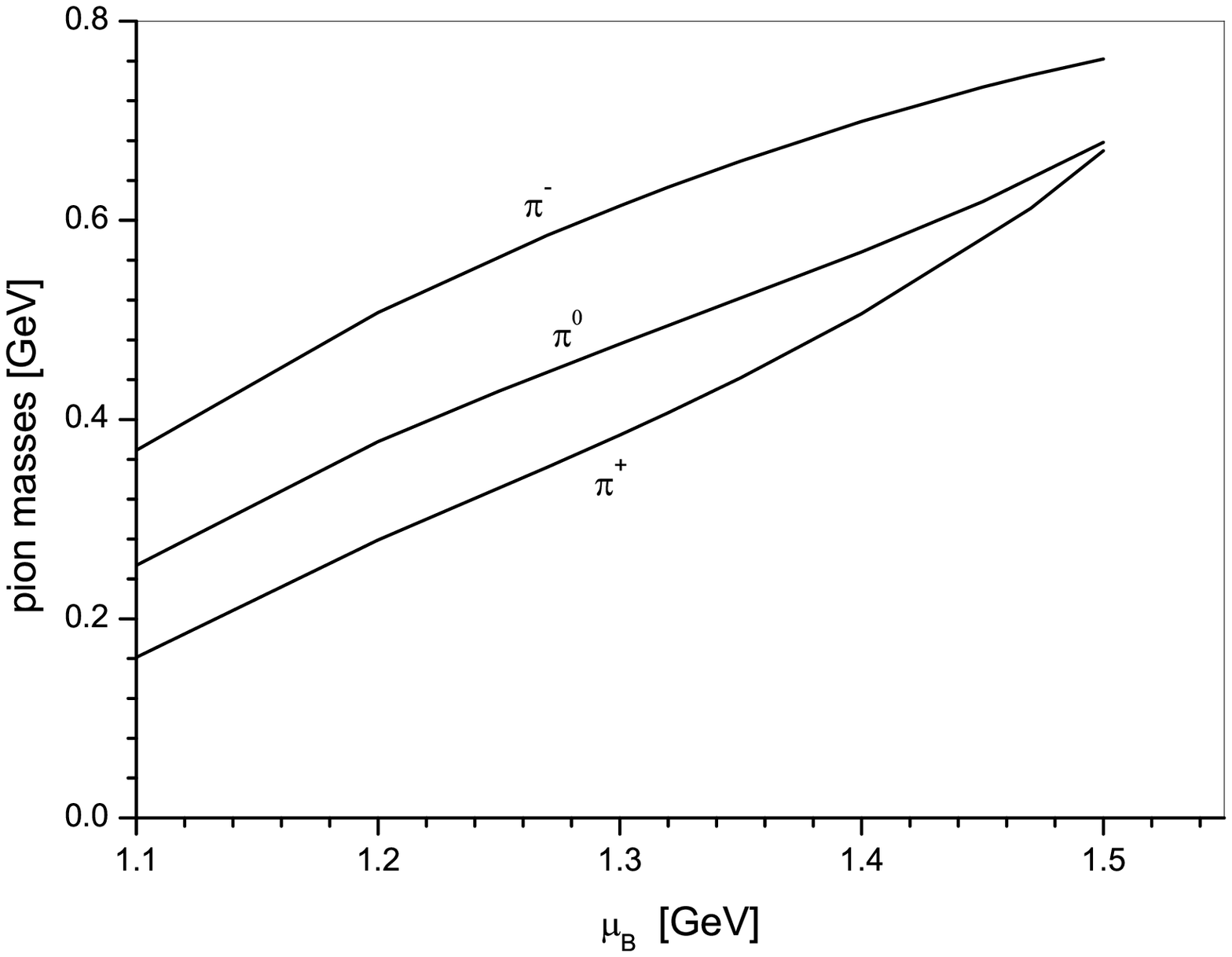}
\hfill
\includegraphics[width=0.45\textwidth]{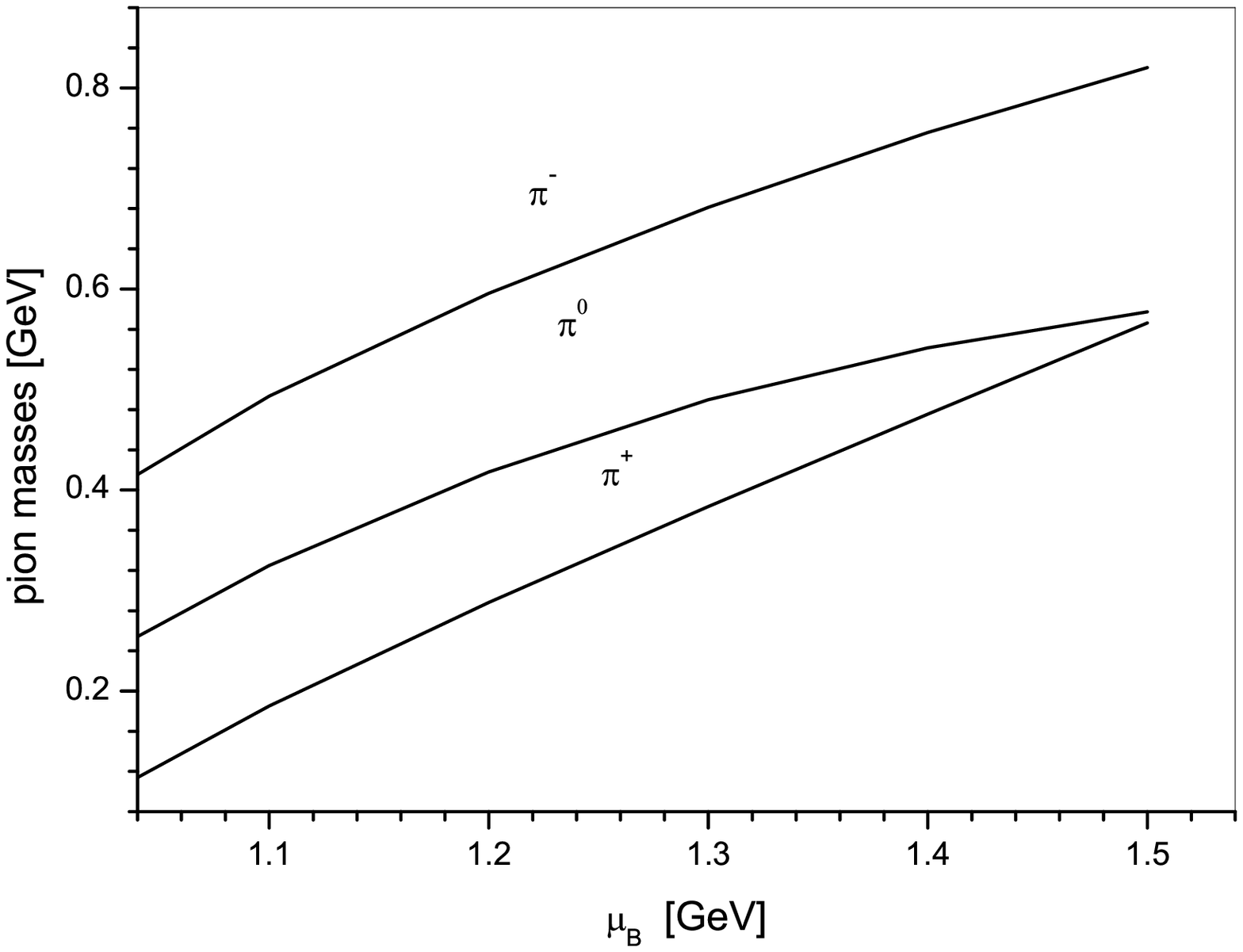}\\
\parbox[t]{0.45\textwidth}{
 \caption{The behaviour of meson masses vs $\mu_B$ in the gapless
 ($H=3G/4$) color superconducting neutral matter
 ($\mu_B>\mu_B^c=1.08$ GeV).}
 \label{fig:5}}\hfill
\parbox[t]{0.45\textwidth}{\caption{The behaviour of meson masses vs
$\mu_B$ in the gapped
($H=G$) color superconducting neutral matter ($\mu_B>\mu_B^c=1.04$
GeV).} \label{fig:6} }
\end{figure}

\subsection{The $\pi^0,\sigma$-meson masses}
\label{pi0}

As it was already discussed after (\ref{35}), $\pi^0$-mesons are not
mixed with other fields in the framework of the NJL model (\ref{3}),
i.~e.\  the 1PI Green functions of the form $\Gamma_{\pi^0 X}(x-y)$,
where $X(x)=\sigma(x),\pi^\pm(x),\Delta_A(x),\Delta^*_{A'}(x)$, are
equal to zero.  In this case the 1PI Green function
$\Gamma_{\pi^0\pi^0}(x-y)$ is just the inverse $\pi^0$-propagator,
which can be found from (\ref{34}). To get the $\pi^0$-mass we
need the expression $\overline{\Gamma_{\pi^0\pi^0}}(p)$, i.~e.\  the
Fourier transform of $\Gamma_{\pi^0\pi^0}(x-y)$, in the rest frame.
Using the technique presented in the previous section, one can obtain
after tedious but straightforward calculations at $p=(p_0,0,0,0)$:
\begin{eqnarray}
&&~~~~~~~~\overline{\Gamma_{\pi^0\pi^0}}(p_0)=\frac{1}{2G}
+8\int\frac{d^3q}{(2\pi)^3}
\frac{E_\Delta^+E_\Delta^-+E^+E^-+\Delta^2}{E_\Delta^+E_\Delta^-}
\frac{E_\Delta^++E_\Delta^-}{p_0^2-(E_\Delta^++E_\Delta^-)^2}+
\nonumber\\&&
4\int\frac{d^3q}{(2\pi)^3}\left\{\frac{\theta(\delta\mu-
E_\Delta^-)+\theta(-\delta\mu-E_\Delta^-)}{E_\Delta^-}\left[
\frac{(p_0+E_\Delta^-)E_\Delta^--E^+E^--|\Delta|^2}
{(p_0+E_\Delta^-)^2-(E_\Delta^+)^2}+
\frac{(E_\Delta^--p_0)E_\Delta^--E^+E^--|\Delta|^2}
{(E_\Delta^--p_0)^2-(E_\Delta^+)^2}\right]
\right.\nonumber\\
&&\left.+\frac{\theta(\delta\mu-E_\Delta^+)+\theta(-\delta\mu-
E_\Delta^+)}{E_\Delta^+}\left[
\frac{(E_\Delta^+-p_0)E_\Delta^+-E^+E^--|\Delta|^2}
{(E_\Delta^+-p_0)^2-(E_\Delta^-)^2}+
\frac{(E_\Delta^++p_0)E_\Delta^+-E^+E^--|\Delta|^2}
{(E_\Delta^++p_0)^2-(E_\Delta^-)^2}\right]\right\}\nonumber\\
&&+8\int\frac{d^3q}{(2\pi)^3}\frac{E}{p_0^2-4E^2}\left[
\theta(E-\mu_{db})+
\theta(E-\mu_{ub})-
\theta(-\mu_{ub}-E)-\theta(-\mu_{db}-E)
\right].
\label{41}
\end{eqnarray}
The function (\ref{41}) is an even one, i.~e.\
$\overline{\Gamma_{\pi^0\pi^0}}(p_0)=$$\overline{\Gamma_{\pi^0\pi^0}}
(-p_0)$. Hence, it effectively depends on the variable $p_0^2$.
Numerical investigations, performed both at $H=3G/4$ and $H=G$, show
that for each fixed value of $\mu_B>\mu_B^c$ the 1PI Green function
(\ref{41}) has a single zero on the positive $p_0^2$-semi-axis, which
is just the mass squared of the $\pi^0$-meson. Its mass in the 2SC
phase of the dense and neutral matter is presented graphically in
Figs 5,6.

The situation with the $\sigma$-meson mass is much more involved.
Indeed, as it follows from the previous section, in the model
under consideration the $\sigma$-meson is mixed with
$\Delta_2(x),\Delta^*_{2} (x)$-diquarks (see also
\cite{bekvy,eky1,eky2}). So to get the particle masses in this
case, one should find the zeros of the 3$\times$3-matrix
determinant, whose matrix elements are nothing but the 1PI Green
functions $\overline{\Gamma_{XY}}(p_0)$ (we use the rest frame in
the momentum space representation), where
$X(x),Y(x)=\sigma(x),\Delta_2(x), \Delta^*_{2}(x)$. In general, it
is a rather hard task, which however can be significantly
simplified due to some reasons. It turns out that 1PI Green
functions of the form  $\Gamma_{\sigma X}$
($X(x)=\Delta_2(x),\Delta^*_{2}(x)$) are proportional to the gap
$\Delta$ (see Fig. 2) and constituent quark mass $M$ (see Fig.1)
as well. Hence, in the normal color symmetric phase (where
$\Delta=0$) there is no mixing between the $\sigma$-meson and
diquarks $\Delta_2(x), \Delta^*_{2}(x)$ at all. The similar is
true for the 2SC phase, if the current quark mass $m$ is zero (in
this case, as was pointed out in the footnote \ref{f4}, the
parameter $M$ is also equal to zero in the 2SC phase). In our
consideration the parameter $M$ is a rather small quantity in
comparison to the gap $\Delta$ in the 2SC phase (see Figs 1,2). So
to have a grasp of the order of magnitude of the $\sigma$-meson
mass, we ignore for simplicity the $\sigma-\Delta_2$ mixing effect
in this case, too. As a result, we see that both for $H=3G/4$ and
$H=G$ the $\sigma$-meson mass is defined by the 1PI Green function
$\Gamma_{\sigma\sigma}(x-y)$ which is approximately the 1PI Green
function $\Gamma_{\pi^0\pi^0}(x-y)$. Hence, for both above
mentioned values of the coupling constant $H$ the $\sigma$-meson
mass is approximately equal to the $\pi^0$-meson mass in the 2SC
neutral matter (see Figs 5,6).

~

~ Now some comments, concerning the behaviour of the $\pi$-meson
masses in the neutral dense matter, are in order. First of all,
note that in the normal phase (where $\mu_B<\mu_B^c$) the
$\mu_Q$-term of the Lagrangian (\ref{3}) is zero. As a result, the
ground state of this phase is an SU(2)$_I$-invariant one. Due to
this symmetry, in the normal phase all $\pi$-mesons have a common
mass that is approximately 140 MeV for all $\mu_B<\mu_B^c$ both
for $H=3G/4$ and $H=G$ (the $\sigma$-meson mass is approximately
equal to 700 MeV in the normal phase). Above the critical point,
i.~e.\  at $\mu_B>\mu_B^c$, the isotopic SU(2)$_I$ symmetry of the
system is broken due to the appearance in (\ref{3}) of a nonzero
$\mu_Q$-term. So in the 2SC phase of electrically neutral matter
the meson masses are allowed to have different values. Just this
conclusion was supported by our numerical investigations (see
Figs.\ 5,6), where a rather strong splitting of $\pi$-meson masses
is observed. In contrast, if the electric charge neutrality is not
imposed on the NJL-system, then all mesons have a common mass in
the 2SC phase \cite{eky1,eky2}. 

Finally, note that in \cite{Ebert:2005wr} a rather strong splitting
of $\pi$-meson masses was shown to exist in electrically neutral and
noncolor--superconducting dense quark matter both with or without
pion condensation phenomenon.

\section{Diquark masses}

As in the previous section, we will ignore, for simplicity, the
mixture between diquarks $\Delta_2(x),\Delta^*_{2}(x)$ and
$\sigma$-meson (this is justified by moderately small values of the
parameter $M$ in the 2SC phase). In this case, in order to obtain the
masses of diquarks, we need to analyze the 1PI Green functions,
generated by the effective action ${\cal S}^{(2)}_{\rm diquarks}$
(\ref{25}):
\begin{eqnarray}
&& \Gamma_{XY}(x-y)=-\frac{\delta^2{\cal S}^{(2)}_{\rm
diquarks}}{\delta Y(y)\delta X(x)},
 \label{42}
\end{eqnarray}
where $X(x),Y(x)=\Delta_A(x),\Delta^*_{A'}(x)$.

\subsection{Diquark masses in the 2SC phase}

Due to the structure
of (\ref{25}), the diquarks, as such, are not mixed to one another in
the framework of our model. So it is reasonable to study step-by-step
the diquark excitations of the 2SC neutral matter in the
$\Delta_2(x),\Delta^*_{2}(x)$-, $\Delta_5(x),\Delta^*_{5}(x)$-, and
finally in the  $\Delta_7(x),\Delta^*_{7}(x)$ sectors of the model.

The investigations of the 1PI Green functions (\ref{42}) in the
$\Delta_5(x)$ and $\Delta_{7}(x)$ sectors of the model supply us with
four excitations of the 2SC phase ground state. All of them have
the common mass $3|\mu_8|$ both for $H=3G/4$ and $H=G$, when color-
and electric charge neutrality constraints are imposed
\cite{he,eky2}. Evidently, these excitations form two real (or
one complex) doublets of the SU(2)$_c$ ground state symmetry group of
the 2SC phase.

To study the masses of the SU(2)$_c$-singlet diquark excitations, we
should consider the 1PI Green functions in the
$\Delta_2(x),\Delta^*_{2}(x)$ sector of the model. It can be shown in
the usual way that in the 2SC phase these quantities take the
following form (here again the rest frame, i.~e.\  $\vec
p=0$, in the momentum space representation is used):
\begin{eqnarray}
\overline{\Gamma_{\Delta_2\Delta_2}}(p_0)=\overline{\Gamma
_{\Delta^{*}_2\Delta^{*}_2}}(p_0)=
4\Delta^2I_0(p_0^2),~~~~~~~\overline{\Gamma_{\Delta^{*}_2
\Delta_2}}(p_0)
=\overline{\Gamma_{\Delta_2\Delta^{*}_2}}(-p_0)=
(4\Delta^2-2p_0^2)I_0(p_0^2)+4p_0I_1(p_0^2),
\label{43}
\end{eqnarray}
where
\begin{eqnarray}
&&I_0(p_0^2)=\int\frac{d^3q}{(2\pi)^3}
\frac{\theta(E_\Delta^+-|\delta\mu|)}{E_\Delta^+[4(E_\Delta^+)^2-
p_0^2]}+\int\frac{d^3q}{(2\pi)^3}
\frac{\theta(E_\Delta^--|\delta\mu|)}{E_\Delta^-[4(E_\Delta^-)^2-
p_0^2]},\nonumber\\
&&I_1(p_0^2)=\int\frac{d^3q}{(2\pi)^3}
\frac{\theta(E_\Delta^+-|\delta\mu|)E^+}{E_\Delta^+[4(E_\Delta^+)^2-
p_0^2]}-\int\frac{d^3q}{(2\pi)^3}
\frac{\theta(E_\Delta^--|\delta\mu|)E^-}{E_\Delta^-[4(E_\Delta^-)^2-
p_0^2]}.
\label{44}
\end{eqnarray}
From the expressions (\ref{43}) it is possible to compose the inverse
propagator matrix $\mathcal{G}^{-1}(p_0)$ for the diquarks
$\Delta_2(x),\Delta^*_{2}(x)$ moving in neutral 2SC matter:
\begin{equation}
  \label{45}
\mathcal{G}^{-1}(p_0)=-  \left(\begin{array}[c]{cc}
\overline{\Gamma_{\Delta_2\Delta_2}}(p_0),&
\overline{\Gamma_{\Delta_2\Delta^*_2}}(p_0)\\
\overline{\Gamma_{\Delta^*_2\Delta_2}}(p_0),&
\overline{\Gamma_{\Delta^*_2\Delta^*_2}}(p_0)
        \end{array}\right).
\end{equation}
Then, the mass spectrum in the $\Delta_2(x),\Delta^*_{2}(x)$ sector
of the model is defined by the equation
\begin{equation}
  \label{46}
{\rm det}\mathcal{G}^{-1}(p_0)\equiv
4p_0^2\big\{(p_0^2-4\Delta^2 )I_0^2(p_0^2)-4I_1^2(p_0^2)\big\}\equiv
4p_0^2F(p_0^2)=0.
\end{equation}
In the $p_0^2$-plane this equation has an evident zero,
corresponding to a Nambu--Goldstone boson (NG),  $p_0^2=0$. (Since
$F(0)\ne 0$, it does not have another zeros of the form
$p_0^2=0$.) \footnote{It might seem that the appearance of the
NG-boson in the mass spectrum of the model is strongly connected
with the disregarding of the mixing between $\sigma$-meson and
$\Delta_2(x),\Delta^*_{2}(x)$ diquarks. However, as it was shown
in \cite{eky1}, this NG-boson is a native property of the model,
since it exists in the mass spectrum even in the case, when the
mixing is taken into account.} Since in the 2SC phase the chemical
potential $\mu_8$ is nonzero, it is clear that in this phase the
initial color SU(2)$_c\times$U(1)$_{\lambda_8}$ symmetry of the
Lagrangian (\ref{3}) is spontaneously broken down to SU(2)$_c$
group. So, there is only one broken symmetry generator,
corresponding to the above mentioned NG-boson solution of the
equation (\ref{46}). Hence, it is possible to assert that in the
2SC phase of color- and electrically neutral matter, described by
the NJL model (\ref{3}), there appears a {\bf normal number} of
NG-bosons, corresponding to a spontaneous breaking of the color
symmetry. In contrast, if neutrality requirements are not imposed
and only the baryon chemical potential is taken into account, then
in the NJL model there is an {\bf abnormal number} of NG-bosons in
the mass spectrum of the 2SC phase \cite{bekvy}. (Note, an
abnormal number of NG-bosons is not a quite unexpected phenomenon.
It is inherent to a variety of quantum models with broken Lorentz
symmetry, which is provided by chemical potentials
\cite{pirner,brauner}.)

For the further investigation of the equation (\ref{46}) some
additional
information about the functions $I_0,I_1$ (\ref{44}), and
consequently about the function $F$ in (\ref{46}), is required.
It turns out that these functions are analytical in the complex
$p_0^2$-plane, except for the cut $\varkappa<p_0^2$ along the real
axis (the first Riemann sheet). It is easy to verify that $\varkappa
=4\Delta^2$ in the case of the gapped 2SC phase, i.~e.\  at $H=G$,
where $\Delta>|\delta\mu|$. However, $\varkappa =4|\delta\mu|^2\equiv
\mu_Q^2>4\Delta^2$ in the case of the gapless 2SC phase, i.~e.\  at
$H=3G/4$, where $\Delta<|\delta\mu|$. The quantity $F(p_0^2)$ is a
complex-valued function in the whole first Riemann sheet, except for
points on the real axis which do not belong to a cut, where
$F(p_0^2)$ is a real-valued function.

Notice that in the case of the gapless 2SC phase, the cut of the
$p_0^2$-plane originates to the right of the point $4\Delta^2$,
whereas in the case of the gapped 2SC phase it just starts in the
point $4\Delta^2$. This circumstance is of decisive importance for
the appearance of the diquark mass difference in the two types of
color superconductivity. Indeed, at $H=G$ we did not manage to
find any solution of the equation $F(p_0^2)=0$ in the first
Riemann sheet of the variable $p_0^2$. (In particular, it is
evident from (\ref{46}) that in the real points such that
$p_0^2<4\Delta^2$ the function $F(p_0^2)$ is an exactly negative
quantity.) So using the procedure, presented in the appendix of
our previous paper \cite{eky1}, we continue the function
$F(p_0^2)$ to the second Riemann sheet, where it takes a zero
value in some complex point. It means that an SU(2)$_c$-singlet
diquark resonance appeared in the mass spectrum of the neutral
gapped 2SC phase. Its mass $M_D$ and width $\Gamma$ are presented
in Fig. 7 as functions of $\mu_B$. For the gapless 2SC phase (at
$H=3G/4$) the situation is quite different. In this case the first
Riemann sheet of the function $F(p_0^2)$ contains in addition the
set of real points $p_0^2$ such that $4\Delta^2\leq
p_0^2<\mu_Q^2$. Just among these points the zero $M_D^2$ of
$F(p_0^2)$ is located. It means that the existence of an
SU(2)$_c$-singlet stable diquark excitation with mass $M_D$ such
that  $2\Delta\leq M_D<|\mu_Q|$  (see Fig. 8) is typical for the
neutral gapless 2SC phase (at $H=3G/4$), in contrast to the
neutral gapped one (at $H=G$), where it is a resonance (see Fig.
7). In addition, it is easily concluded that for the same relation
$H=3G/4$ the SU(2)$_c$-singlet diquark is a heavy resonance in the
2SC (gapped) phase without neutrality requirement
\cite{eky1,eky2}, whereas it is a stable particle in the gapless
neutral 2SC phase, if the neutrality constraints are fulfilled.
\footnote{Let us quote also another less rigorous, but
more physical argument in favour of the stability of the
SU(2)$_c$-singlet diquark excitation in the g2SC phase ($H=3G/4$).
Namely, the decay of the above mentioned diquark
excitation with mass $M_D$ into a pair of $u$- and $d$ quark
quasiparticles is forbidden due to the energy disbalance. Indeed, the
total energy $E(p)$ of this quark pair in their center-of-mass system
takes the form $E(p)=E_u(p)+E_d(p)$, where $p=|\vec p|\ge 0$ and
$E_u(p)=E_\Delta^-+|\delta\mu|$, $E_d(p)=|E_\Delta^-- |\delta\mu||$.
One can easily show that $E(p)\ge |\mu_Q|$. Since the mass $M_D$
of the SU(2)$_c$-singlet diquark excitation in the g2SC phase is
smaller than $|\mu_Q|$, we conclude that the decay of this
excitation into a pair of $u$- and $d$ quasiparticles is forbidden.}
\begin{figure}
\includegraphics[width=0.45\textwidth]{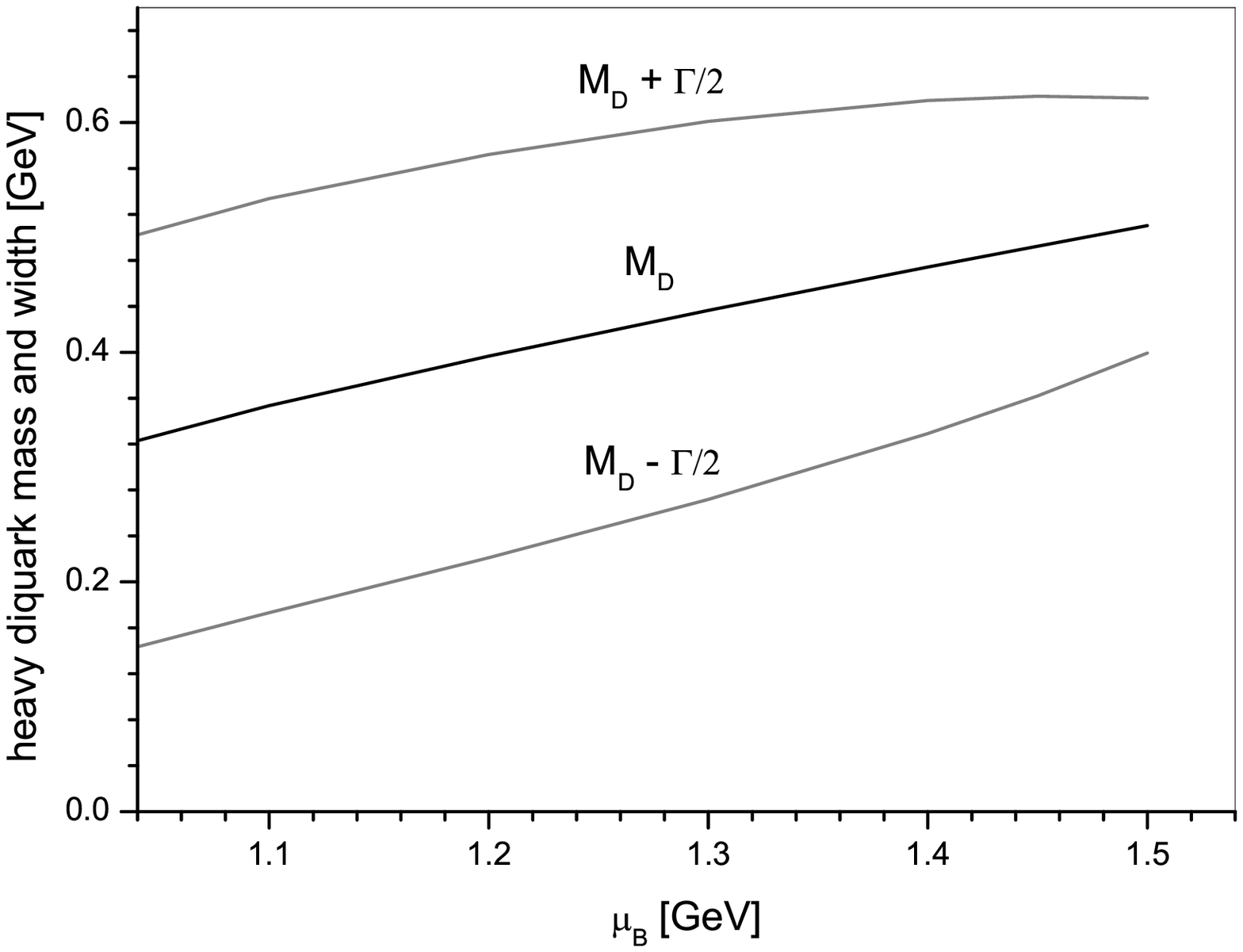}
\hfill
\includegraphics[width=0.45\textwidth]{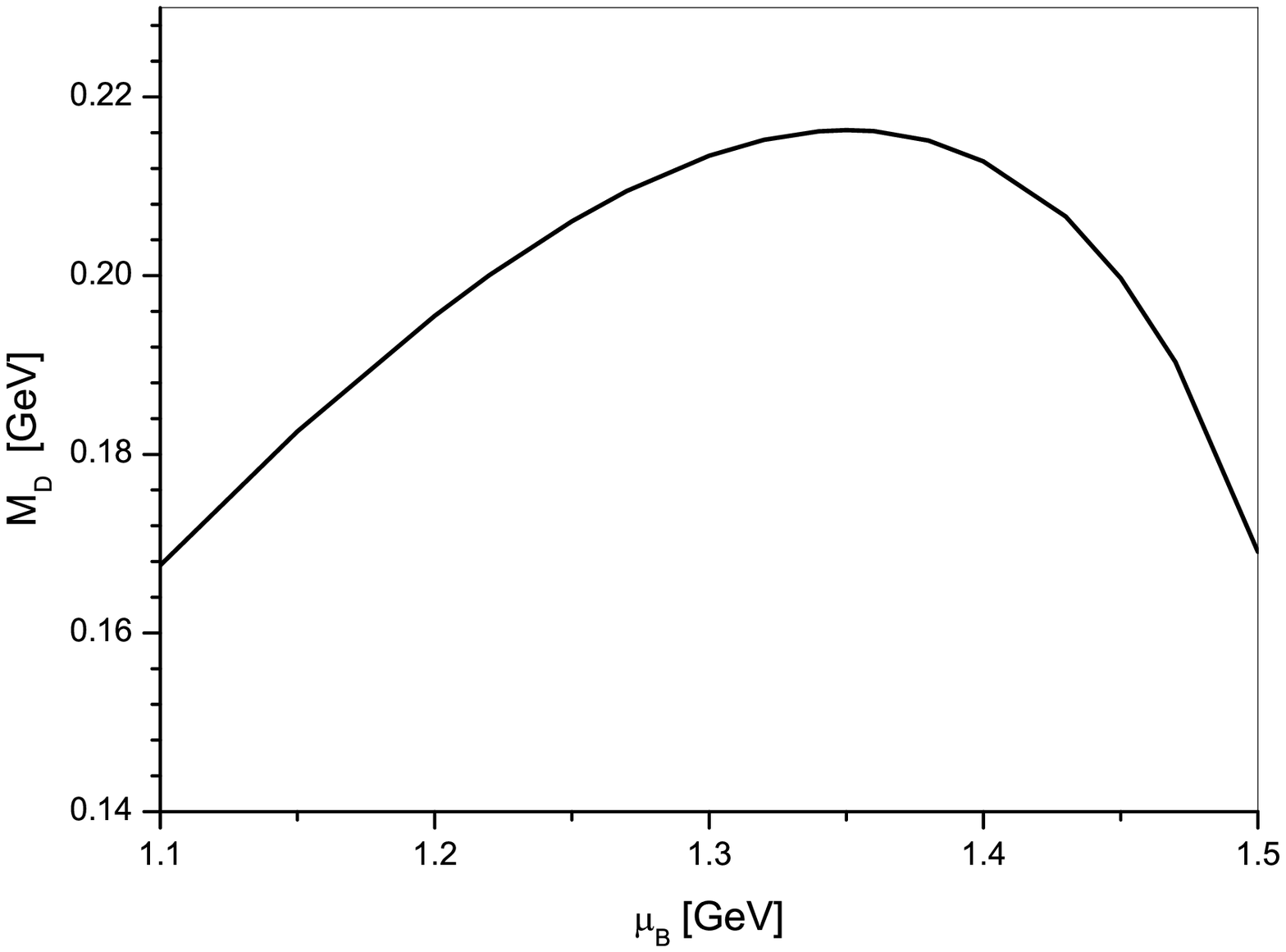}\\
\parbox[t]{0.45\textwidth}{
 \caption{The sketch of the SU(2)$_c$-singlet diquark resonance in
 the
 gapped ($H=G$) 2SC neutral matter ($\mu_B>\mu_B^c=1.04$
GeV). The solid line is for its mass $M_D$ vs $\mu_B$, whereas the
width of the strip between dashed lines $M_D\pm\Gamma/2$ is its width
$\Gamma$ vs $\mu_B$.}
 \label{fig:7}
}\hfill
\parbox[t]{0.45\textwidth}{
\caption{The behaviour of the stable SU(2)$_c$-singlet diquark
mass $M_D$ vs $\mu_B$ in the gapless ($H=3G/4$) color
superconducting neutral matter ($\mu_B>\mu_B^c=1.08$ GeV).}
\label{fig:8} }
\end{figure}

Notice that in the present paper actually the dispersion relations of
mesons and diquarks, i. e. the momentum dependence of their energies,
are investigated for the particular value of the three-momentum,
$\vec p=0$. Contrary, in the recent papers \cite{h,g} the diquark
dispersion relations were studied at $\vec p\ne 0$, and a conclusion
about the Higgs instability of the NJL gapless 2SC phase was made. In
this case the g2SC phase is unstable vs fluctuations of the Nambu --
Goldstone diquark fields and, as a result some of the diquarks
acquire a non-physical negative velocity squared \cite{h}. (In
addition, the g2SC phase of a gauged NJL model suffers also from a
chromomagnetic instability \cite{hashimoto}.) In particular, it was
shown that a sufficient condition for the Higgs instability in the
gapped 2SC phase is the relation $\sqrt{2}|\delta\mu|>\Delta$. Since
in the case $G=H$ this relation is not fulfilled, one may conclude
that in the gapped neutral 2SC phase considered above the Higgs
instability is absent. 

Finally, it is necessary to note that in the physical
SU(2)$_c$-singlet diquark channel, we have found a stable massive
mode in the case of the g2SC phase ($H=3G/4$), whereas in \cite{h} it
was claimed that a gapless tachyon emerges in this channel. Actually,
there is no contradiction between our result and that paper. Indeed,
in our case the diquark mass $M_D$ obeys the relation $2\Delta <
M_D=p_0$. In contrast, in \cite{h} those solutions for the diquark
dispersion relation were investigated that are constrained by $|p_0|,
|\vec p|\ll \Delta$. It means that in the gapless color
superconductor there are two branches for the dispersion relation of
the physical SU(2)$_c$-singlet diquark mode at small values of $|\vec
p|$. The first is the massive heavy excitation with $2\Delta <p_0$
(see Fig. 8 of the present paper), the other one corresponds to a
gapless tachyon, located in a quite different kinematic region with
$|p_0|\ll \Delta$. In addition, it was shown in \cite{h} that the
singularity $p_0/|\vec p|$ appears in the diquark two-point 1PI Green
function in the gapless 2SC phase. In this case, owing to this
singularity, it is quite possible to get at $|\vec p|\to 0$ a result
(gapless tachyon) that does not coincide with the diquark mass,
directly calculated at $\vec p=0$. 
\label{VA}

\subsection{Diquarks in the normal phase ($\Delta =0$,
$\mu_8=0$, $\mu_Q=0$)}\label{VB}

At $\Delta=0$ the three complex diquark fields $\Delta_A(x)$
($A=2,5,7$) are not mixed with other fields in the second order
effective action (\ref{23}) of the model. Since in addition the
quantities $\mu_8$, $\mu_Q$ are equal to zero  in the normal phase,
there is an SU(3)$_c$ symmetry of the ground state of this phase. So,
in order to study the diquark masses at $\mu_B<\mu_B^c$, it is enough
to consider, e.g., the $\Delta_2$-diquark sector only.  In
this phase the inverse propagator matrix $\mathcal{G}^{-1}(p_0)$ for
the diquarks $\Delta_2(x),\Delta^*_{2}(x)$ looks, in contrast to the
inverse propagator (\ref{45}) of these fields in the 2SC phase,
more simpler \cite{eky1,eky2} (again, we use the rest frame, $\vec p
=0$, in the momentum space representation):
\begin{equation}
 \mathcal{G}^{-1}(p_0)=-  \left(\begin{array}[c]{cc}
~~~~~0~~~~~~,&
\overline{\Gamma_{\Delta_2\Delta^*_2}}(p_0)\\
\overline{\Gamma_{\Delta^*_2\Delta_2}}(p_0),&
0        \end{array}\right),
\label{47}
\end{equation}
where $\overline{\Gamma_{\Delta_2\Delta^*_2}}(p_0)=$$\overline{
\Gamma_{\Delta^*_2\Delta_2}}(-p_0)$,
\begin{eqnarray}
\overline{\Gamma_{\Delta^*_2\Delta_2}}(p_0)=
\frac 1{4H}-16
\int\frac{d^3q}{(2\pi)^3}\frac{E}{4E^2-(p_0+2\mu_B/3)^2}\equiv
\frac 1{4H}-\Phi(\epsilon),
\label{48}
\end{eqnarray}
and $\epsilon=(p_0+2\mu_B/3)^2$. Since the determinant of the
inverse propagator matrix (\ref{47}) takes the form
\begin{eqnarray}
{\rm det}\mathcal{G}^{-1}(p_0)\equiv
\overline{\Gamma_{\Delta^{*}_2\Delta_2}}(p_0)
\overline{\Gamma_{\Delta_2\Delta^{*}_2}}(p_0)=
\overline{\Gamma_{\Delta^{*}_2\Delta_2}}(p_0)
\overline{\Gamma_{\Delta^{*}_2\Delta_2}}(-p_0),
\label{49}
\end{eqnarray}
it is clear that the mass spectrum is defined by the zeros of the 1PI
Green function (\ref{48}).
Note, the function $\Phi(\epsilon)$ is analytical in
the whole complex $\epsilon$-plane, except for the cut
$4M^2<\epsilon$ along the real axis. (In general, this function is
defined on a complex Riemann surface which is to
be described by several sheets. The integral representation for
$\Phi(\epsilon)$, given in (\ref{48}), defines its values on the
first sheet only. To find a value of $\Phi(\epsilon)$ on the rest of
the Riemann surface, a special procedure of analytical continuation
is needed (see, e.g., in \cite{eky1}).) It turns out that for the
model parameter set (\ref{32}) and a wide set of the coupling
constant $H$ values (see below), the equation
\begin{eqnarray}
\label{50}
\Phi(\epsilon)=\frac 1{4H}
\end{eqnarray}
has on the first Riemann sheet of the variable $\epsilon$ a single
root $\epsilon_0$ on the real axis such that $0<\epsilon_0<4M^2$.
In this case the physical meaning of $\epsilon_0$ is that
$\epsilon_0=(M_D^o)^2$, where $M_D^o$ is the mass of a stable diquark
at $\mu_B=0$. Having a root $\epsilon_0$, one can find two zeros of
the 1PI Green function
$\overline{\Gamma_{\Delta^{*}_2\Delta_2}}(p_0)$ as well as four zeros
of the determinant (\ref{49}). They provide us with the following
mass squared in the $\Delta^*_2, \Delta_2$ sector of the model:
\begin{equation}
  \label{51}
(M_{\Delta})^2=(M_D^o-2\mu_B/3)^2,~~~~~(M_{\Delta^{*}})^2=
(M_D^o+2\mu_B/3)^2.
\end{equation}
In particular, if $H=3G/4$, then $M_D^o\approx 1.988M$, if $H=G$,
then $M_D^o\approx 1.746M$ (here $M$ is the
constituent quark mass, or gap, in the normal phase (see Fig. 1),
i.~e.\  $M\approx 0.350$ GeV). We relate   $M_{\Delta}$
in (\ref{51}) to the mass of the diquark with the baryon
number $B=2/3$ and $M_{\Delta^{*}}$ to the mass of the antidiquark
with $B=-2/3$. The difference between diquark and antidiquark masses
in (\ref{51}) is explained by the absence of a charge conjugation
symmetry in the presence of a chemical potential $\mu_B$.

Finally,  due to the underlying color SU(3)$_c$ symmetry, the
previous statement is valid also for $\Delta^*_5, \Delta_5$ and
$\Delta^*_7, \Delta_7$. As a result, we have a color antitriplet of
diquarks  with the mass $M_{\Delta}$ (\ref{51})  as well as a
color triplet of antidiquarks with the mass $M_{\Delta^{*}}$ in the
mass spectrum of the normal phase, i.~e.\  at $\mu_B<\mu_B^c$.

It is clear from the equation (\ref{50}) that its own solution
$\epsilon_0$  lies inside the interval $0<\epsilon_0<4M^2$ only if
$H^*<H<H^{**}$, where $H^*$ and $H^{**}$ are defined by
\begin{eqnarray}
H^* &\equiv&\frac {1}{4\Phi(4M^2)}= \frac {\pi^2}{4\left
[\Lambda\sqrt{M^2+\Lambda^2}+M^2\ln((
\Lambda+\sqrt{M^2+\Lambda^2})/M)\right ]},\nonumber\\
H^{**}&\equiv&\frac {1}{4\Phi(0)}=\frac {\pi^2}{4\left
[\Lambda\sqrt{M^2+\Lambda^2}-M^2\ln((
\Lambda+\sqrt{M^2+\Lambda^2})/M)\right ]}=\frac{3MG}{2(M-m)}
\label{52}
\end{eqnarray}
(here the values of $\Lambda$, $m$ and $G$ are presented in
(\ref{32}), whereas $M\approx 0.350$ GeV is the dynamical quark mass
in the normal phase). In this case, as was noted above, $\epsilon_0$
defines the stable diquark mass $M_D^o$ in the vacuum, i.~e.\  at
$\mu_B=0$, through the relation $\epsilon_0=(M_D^o)^2$. The values of
$M_D^o$ are depicted in Fig. 9 as a function of the parameter
$\eta=H/G$. (Note, $\eta^*=H^*/G$,
$\eta^{**}=H^{**}/G=$$1.5M/(M-m)$.) For a rather weak interaction
in the diquark channel ($H<H^*$ or $\eta<\eta^*$),  $\epsilon_0$ runs
onto the second Riemann sheet, and unstable diquark modes
(resonances) appear. Unlike this, a sufficiently strong interaction
in the diquark channel ($H>H^{**}$ or $\eta>\eta^{**}$) pushes
$\epsilon_0$ towards the negative semi-axis, i.~e.\  $(M_D^o)^2<0$ in
this case. The latter indicates a tachyon singularity in the diquark
propagator, evidencing that the SU(3)$_c$-color symmetric ground
state is not stable. Indeed, at a very large  $H$, as it has been
shown in \cite{ek}, the color symmetry is spontaneously broken even
at a vanishing chemical potential. The fact that at
$\eta\to\eta^{**}_-$ the diquark mass $M_D^o$ tends to zero may be
considered as a precursor of the spontaneous breaking of the
SU(3)$_c$ symmetry, taking part at $H=H^{**}$.
\begin{figure}
\includegraphics[width=0.45\textwidth]{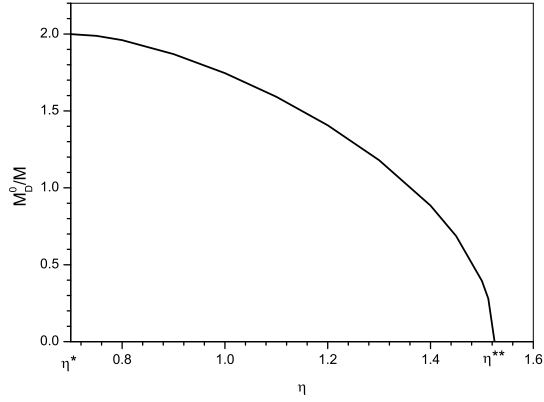}
 \caption{The behavior of the stable diquark mass $M_D^o$  vs
 $\eta\equiv H/G$ in the vacuum (at $\mu_B=0$). Here
 $\eta^*=H^*/G\approx  0.698$, $\eta^{**}=H^{**}/G\approx 1.525$,
 $M\approx 0.350$ GeV. At  $\eta<\eta^*$ there are no stable diquarks
 in the normal phase, and these particles are resonances. At
 $\eta>\eta^{**}$ an SU(3)$_c$  symmetric ground state of the normal
 phase (including the vacuum) is unstable in favor of color
 superconducting phases.}
 \end{figure}

\section{Summary and discussion}

The present paper is the last in the series of our papers
\cite{bekvy,eky1,eky2}, devoted to the investigation of the ground
state bosonic excitations (mesons and diquarks) of color
superconducting quark matter. The novel features of our present
consideration, performed in the framework of the two-flavored NJL
model, are the local electric charge neutrality constraint as well
as the $\beta$-equilibrium that is due to taking into account
electrons. As a result, depending on the diquark channel coupling
constant $H$, the gapless ($H=3G/4$) or gapped ($H=G$) 2SC may
exist in the system based on the Lagrangian (\ref{3}). Since in
(\ref{3}) a new term with the electric charge chemical potential
$\mu_Q$ appeared, we have an explicit breaking of the flavor SU(2)
symmetry in both 2SC phases. So, in contrast to the non-neutral
case \cite{bekvy,eky1,eky2}, $\pi$-mesons acquire a rather strong
mass splitting in both gapless- and gapped 2SC phases (see Figs.\
5,6).

The diquark sector of the model consists of six modes. Since in the
color and electrically  neutral 2SC phase the chemical potential
$\mu_8$ is not equal to zero (see Fig. 3), the Lagrangian (\ref{3})
is invariant under the color SU(2)$_c\times$U(1)$_{\lambda_8}$ group.
It turns out that in the color superconducting phase this symmetry is
spontaneously broken down to SU(2)$_c$, so, in accordance with
general theorems, one of the diquark modes takes zero mass, i.~e.\
it is the Nambu -- Goldstone boson that is an SU(2)$_c$-singlet.
Besides, in this phase there are four very light diquark excitations
that are composed into two real SU(2)$_c$-doublets with a common mass
proportional to $\mu_8$. In our opinion, when the NJL model is
gauged, the above mentioned diquark modes should be absent in the
model. Instead, due to the Anderson -- Higgs mechanism, five
massive gluons will appear. So we believe that five  diquark
excitations with zero or very small mass, observed in the 2SC phase
of the nongauged model (\ref{3}), are not physically interesting
objects and might be ignored (see also the discussion in
\cite{hejz}).

The remaining one, an SU(2)$_c$-singlet diquark excitation of the
2SC ground state, has different properties under different external
conditions. Indeed, if the electric charge neutrality requirement is
not imposed, then, as it was shown in \cite{eky1,eky2} at $H=3G/4$,
this diquark mode is a heavy resonance with mass approximately equal
to 1100 MeV. However, if the neutrality and $\beta$-equilibrium
conditions are imposed, then, on the one hand, at $H=3G/4$ the
gapless 2SC phase is realized, in which the above mentioned diquark
mode is already a stable particle, whose mass is evaluated around 200
MeV (see Fig. 8). Note that its mass is approximately equal to the
value of $|\mu_Q|$ presented by Fig. 4. On the other hand, at $H=G$
we have the neutral gapped 2SC phase, in which the diquark is still a
resonance but with a much more smaller mass (see Fig. 7).
 
It was shown, e.g., in \cite{pinto} that at sufficiently high baryon
densities, comparable with densities inside neutron stars, the normal
quark matter equation of state is significantly influenced by scalar
mesons ($\sigma$-meson etc). Moreover, these mesons might change
significantly the mass and radius of the neutron star as well as the
role of hyperon degrees of freedom in dense matter. Since the masses
of diquarks in the electrically neutral color superconducting quark
matter are of the same order as those of mesons (compare Figs 5,6 and
Figs 7,8), one might expect similar effects in the measurable
parameters of neutron stars (e.g. their masses, radii etc), when
diquarks will be taken into account properly in neutron star physics
(see also \cite{pirner}). So, the influence of rather light diquark
excitations might in principle be checked by astrophysical
observations.

In fact, in the present paper the dispersion relations for mesons and
diquarks following from the NJL model (\ref{3}) are investigated in
the rest frame, i.e. at $|\vec p|=0$. In the recent paper \cite{h}
(see also  \cite{g}) the diquark dispersion relations at small
nonzero $|\vec p|$ were studied in the same model. It was found there
that in the gapless 2SC phase ($H=3G/4$) Nambu -- Goldstone diquark
modes have a non-physical negative velocity squared, i.e. the model
is unstable against fluctuations of these fields. (In contrast, as it
follows from our discussion at the end of section \ref{VA}, in
the gapped 2SC phase ($H=G$) this kind of instability is absent.)
Moreover, it was proved in \cite{h} that at small $|\vec p|\ne 0$ in
the SU(2)$_c$-singlet diquark channel of the g2SC phase there is a
gapless tachyon with $|p_0|\ll\Delta$. Our result, i. e.
the existence of a stable diquark excitation with mass $M_D\sim$ 200
MeV (see Fig. 8) in the same channel, does not conflict with this, 
since $p_0=M_D$ belongs to a quite another kinematic region, where
$p_0=M_D>2\Delta$. In spite of the fact that in our calculations
$|\vec p|=0$, we believe that at $|\vec p|\ne 0$ there is also a
branch of the SU(2)$_c$-singlet diquark dispersion relation, which
corresponds to a stable excitation of the g2SC phase with mass $M_D$
(see Fig. 8).    

Moreover, in the present paper the behaviour of the diquark mass at
vanishing $\mu_B$ as a function of the coupling constant $H$ is also
obtained (see Fig. 9).

\section*{Acknowledgments}

Two of the authors (K. G. K. and V. L. Yu.) acknowledge the kind
hospitality of the colleagues of the Particle Theory Group at the
Humboldt University, Berlin. D.E. is grateful to useful discussions
with M. K. Volkov and to the kind hospitality of the colleagues in
the Bogoliubov Laboratory for Theoretical Physics of JINR Dubna.
This work has been supported in part by DFG-project 436 RUS
113/477/0-2, RFBR grant 05-02-16699 and the grant of the
Heisenberg-Landau project. The work by V. L. Yu. was also supported
in part by RFBR Grant 05-02-17695 and by a special program of the
Ministry of Education and Science of the Russian Federation, Grant
RNP.2.1.1.5409.

\appendix

\section{Some formulae}
\label{ApB}
Appendix A contains some useful formulae employed in the text.\\
\noindent
{\bf i) Determinant:}
\begin{eqnarray}
\det\left
(\begin{array}{cc}
A~, & B\\
C~, & D
\end{array}\right )=\det [-CB+CAC^{-1}D]=\det
[DA-DBD^{-1}C].
\label{B1}
\end{eqnarray}
{\bf ii) Inverse matrix:}
\begin{eqnarray}
\left (\begin{array}{cc}
A~, & B\\
C~, & D
\end{array}\right )^{-1}=\left (\begin{array}{cc}
C^{-1}DL~, & -N\\
-L~~,&\!\!\!\!\!\! B^{-1}AN
\end{array}\right )=\left (\begin{array}{cc}
\bar L~~~~, &\!\!\!\!\!\! -A^{-1}B\bar N\\
-D^{-1}C\bar L~, & \bar N
\end{array}\right ),
\label{B2}
\end{eqnarray}
where
\begin{eqnarray}
L=[AC^{-1}D -B]^{-1}~~,~~N=[DB^{-1}A -C]^{-1}~~,
~~\bar L=[A-BD^{-1}C]^{-1}~~,~~\bar N=[D-CA^{-1}B]^{-1}.
\label{B3}
\end{eqnarray}
{\bf iii) Variational derivatives:} Let $A,B$ are some operators in
the coordinate space with matrix elements $A(x,y)\equiv$$A(x-y)$ and
$B(x,y)\equiv B(x-y)$, respectively. Moreover, let $\sigma(x)$ and
$\phi (x)$ are some fields. Then,
\begin{eqnarray}
{\rm Tr}\{A\sigma B\phi\}\equiv\int dx dy dz du
A(x,z)\sigma(z)\delta(z-y)B(y,u)\phi (u)\delta(u-x)=
\int dxdy A(x,y)\sigma(y)B(y,x)\phi(x).
\label{B4}
\end{eqnarray}
It follows from (\ref{B4}) that
\begin{eqnarray}
\frac{\delta^2{\rm Tr}\{A\sigma
B\phi\}}{\delta\sigma(y)\delta\phi(x)}=
A(x,y)B(y,x)=A(x-y)B(y-x).
\label{B5}
\end{eqnarray}

\section{Quark propagator in the Nambu--Gorkov representation}
\label{ApA}

In the Nambu -- Gorkov representation the inverse quark propagator
matrix $S_0^{-1}$ is defined in (\ref{19}). Using the relation
(\ref{B2})  as well as the energy projection operator technique of
\cite{hzc}, one can obtain the following
expressions for the matrix elements $S_{ij}(x-y)$ of the quark
propagator $S_0(z)$ (here $z=x-y$):
\begin{eqnarray}
&&S_{11}(z)=A_{11}(z)\tau_+\tau_-+B_{11}(z)\tau_-\tau_+;~~~S_{12}(z)=
A_{12}(z)\tau_++B_{12}(z)\tau_-;\nonumber\\&&S_{22}(z)=A_{22}(z)
\tau_+\tau_-+B_{22}(z)\tau_-\tau_+; ~~~S_{21}(z)=
A_{21}(z)\tau_++B_{21}(z)\tau_-,
 \label{A1}
\end{eqnarray}
where their explicit structure in the flavor space is presented with
the help of $\tau_\pm\equiv (\tau_1\pm\tau_2)/\sqrt{2}$ matrices
(recall, $\tau_i$ are the Pauli matrices) and
\begin{eqnarray}
&&A_{11}(z)=\frac 12\int\!\frac{d^4q}{(2\pi)^4}e^{-iqz}
\left\{\frac{(q_0+\delta\mu)-E^+}{(q_0+\delta\mu)^2-
(E_\Delta^+)^2}\gamma^0\bar\Lambda_++
\frac{(q_0+\delta\mu)+E^-}{(q_0+\delta\mu)^2-
(E_\Delta^-)^2}\gamma^0\bar\Lambda_-\right\}P_{12}^{(c)}
\nonumber\\
&&~~~~~~~~+\frac 12\int\!\frac{d^4q}{(2\pi)^4}e^{-iqz}
\left\{\frac{\gamma^0\bar\Lambda_+}{q_0+E+\mu_{ub}}+
\frac{\gamma^0\bar\Lambda_-}{q_0+\mu_{ub}-E}
\right\}P_{3}^{(c)}\equiv a_{11}^{(12)}(z)P_{12}^{(c)}+
a_{11}^{(3)}(z)P_{3}^{(c)},
 \label{A2}
\end{eqnarray}
\begin{eqnarray}
&&B_{11}(z)=\frac 12\int\!\frac{d^4q}{(2\pi)^4}e^{-iqz}
\left\{\frac{(q_0-\delta\mu)-E^+}{(q_0-\delta\mu)^2-
(E_\Delta^+)^2}\gamma^0\bar\Lambda_++
\frac{(q_0-\delta\mu)+E^-}{(q_0-\delta\mu)^2-
(E_\Delta^-)^2}\gamma^0\bar\Lambda_-\right\}P_{12}^{(c)}
\nonumber\\
&&~~~~~~~~+\frac 12\int\!\frac{d^4q}{(2\pi)^4}e^{-iqz}
\left\{\frac{\gamma^0\bar\Lambda_+}{q_0+\mu_{db}+E}+
\frac{\gamma^0\bar\Lambda_-}{q_0+\mu_{db}-E}
\right\}P_{3}^{(c)}\equiv b_{11}^{(12)}(z)P_{12}^{(c)}+
b_{11}^{(3)}(z)P_{3}^{(c)},
 \label{A3}
\end{eqnarray}
\begin{eqnarray}
&&A_{22}(z)=\frac 12\int\!\frac{d^4q}{(2\pi)^4}e^{-iqz}
\left\{\frac{(q_0-\delta\mu)-E^-}{(q_0-\delta\mu)^2-
(E_\Delta^-)^2}\gamma^0\bar\Lambda_++
\frac{(q_0-\delta\mu)+E^+}{(q_0-\delta\mu)^2-
(E_\Delta^+)^2}\gamma^0\bar\Lambda_-\right\}P_{12}^{(c)}
\nonumber\\
&&~~~~~~~~+\frac 12\int\!\frac{d^4q}{(2\pi)^4}e^{-iqz}
\left\{\frac{\gamma^0\bar\Lambda_+}{q_0-\mu_{ub}+E}+
\frac{\gamma^0\bar\Lambda_-}{q_0-\mu_{ub}-E}
\right\}P_{3}^{(c)}\equiv a_{22}^{(12)}(z)P_{12}^{(c)}+
a_{22}^{(3)}(z)P_{3}^{(c)},
 \label{A4}
\end{eqnarray}
\begin{eqnarray}
&&B_{22}(z)=\frac 12\int\!\frac{d^4q}{(2\pi)^4}e^{-iqz}
\left\{\frac{(q_0+\delta\mu)-E^-}{(q_0+\delta\mu)^2-
(E_\Delta^-)^2}\gamma^0\bar\Lambda_++
\frac{(q_0+\delta\mu)+E^+}{(q_0+\delta\mu)^2-
(E_\Delta^+)^2}\gamma^0\bar\Lambda_-\right\}P_{12}^{(c)}
\nonumber\\
&&~~~~~~~~+\frac 12\int\!\frac{d^4q}{(2\pi)^4}e^{-iqz}
\left\{\frac{\gamma^0\bar\Lambda_+}{q_0-\mu_{db}+E}+
\frac{\gamma^0\bar\Lambda_-}{q_0-\mu_{db}-E}
\right\}P_{3}^{(c)}\equiv b_{22}^{(12)}(z)P_{12}^{(c)}+
b_{22}^{(3)}(z)P_{3}^{(c)},
 \label{A5}
\end{eqnarray}
\begin{eqnarray}
&&A_{12}(z)=-\frac{\Delta\lambda_2}{\sqrt{2}}
\int\!\frac{d^4q}{(2\pi)^4}e^{-iqz}
\left\{\frac{\gamma^5\bar\Lambda_+}{(q_0+\delta\mu)^2-
(E_\Delta^-)^2}+
\frac{\gamma^5\bar\Lambda_-}{(q_0+\delta\mu)^2-
(E_\Delta^+)^2}\right\}\equiv a_{12}(z)\lambda_2,
\label{A6}\\
&&B_{12}(z)=\frac{\Delta\lambda_2}{\sqrt{2}}
\int\!\frac{d^4q}{(2\pi)^4}e^{-iqz}
\left\{\frac{\gamma^5\bar\Lambda_+}{(q_0-\delta\mu)^2-
(E_\Delta^-)^2}+
\frac{\gamma^5\bar\Lambda_-}{(q_0-\delta\mu)^2-
(E_\Delta^+)^2}\right\}\equiv b_{12}(z)\lambda_2,
\label{A7}
\end{eqnarray}
\begin{eqnarray}
&&B_{21}(z)=\frac{\Delta^*\lambda_2}{\sqrt{2}}
\int\!\frac{d^4q}{(2\pi)^4}e^{-iqz}
\left\{\frac{\gamma^5\bar\Lambda_+}{(q_0+\delta\mu)^2-
(E_\Delta^+)^2}+
\frac{\gamma^5\bar\Lambda_-}{(q_0+\delta\mu)^2-
(E_\Delta^-)^2}\right\}\equiv b_{21}(z)\lambda_2,
\label{A8}\\
&&A_{21}(z)=-\frac{\Delta^*\lambda_2}{\sqrt{2}}
\int\!\frac{d^4q}{(2\pi)^4}e^{-iqz}
\left\{\frac{\gamma^5\bar\Lambda_+}{(q_0-\delta\mu)^2-
(E_\Delta^+)^2}+
\frac{\gamma^5\bar\Lambda_-}{(q_0-\delta\mu)^2-
(E_\Delta^-)^2}\right\}\equiv a_{21}(z)\lambda_2.
\label{A9}
\end{eqnarray}
In the above formulae $P_{12}^{(c)}=$diag$(1,1,0)$,
$P_{3}^{(c)}=$diag$(0,0,1)$ are the projectors on the red-green
and blue subspaces of the color space, correspondingly;
$\lambda_2$ is the Gell-Mann matrix; $\bar\Lambda_\pm=\frac
12(1\pm\frac{\gamma^0(\vec\gamma\vec q-M)}{E})$ are projectors on
the solutions of the Dirac equation with positive/negative energy.
The other notations appearing in (\ref{A2})-(\ref{A9}) are
identical to those of (\ref{27}). Note that in
(\ref{A2})-(\ref{A9}) and similar integrals, containing an
integration over the energy variable, the symbol $q_0$ is a
shorthand notation for $q_0+i\varepsilon\cdot {\rm sign}(q_0)$,
where $\varepsilon\to 0_+$. This prescription correctly implements
the roles of $\mu_B$, $\mu_8$ and $\mu_Q$ as chemical potentials
and preserves the causality of the theory (see, e.g.
\cite{chodos}). Introducing the new projectors $\Lambda_\pm=\frac
12(1\pm\frac{\gamma^0(\vec\gamma\vec q+M)}{E})$, it is very
convenient to use in trace calculations the following relations
\begin{eqnarray}
&&\gamma^5\bar\Lambda_\pm\gamma^5=\Lambda_\pm,~~~
\gamma^0\bar\Lambda_\pm\gamma^0=\Lambda_\mp,~~~
\Lambda_\pm^2=\Lambda_\pm,~~~\Lambda_\pm\Lambda_\mp=0,\nonumber\\&&
{\rm Tr}\Lambda_\pm=2,~~~~~~
{\rm Tr}(\Lambda_\pm\bar\Lambda_\pm)=\frac{2\vec
q^2}{E^2},~~~~~~{\rm Tr}(\Lambda_\pm\bar\Lambda_\mp)=
\frac{2M^2}{E^2}.
 \label{A10}
\end{eqnarray}

\end{document}